\documentclass[preprint]{aastex}

\newcommand{\bomba}{core-S\'ersic}
\newcommand{\Bomba}{Core-S\'ersic}

\newcommand{\kmskpc}{km~s$^{-1}$ kpc$^{-1}$}
\newcommand{\gammap}{\ensuremath{\gamma^{\prime}}}
\newcommand{\chinu}{\ensuremath{\chi^{2}_{\nu}}}
\newcommand{\mbh}{\ensuremath{M_{\rm BH}}}
\newcommand{\mej}{\ensuremath{M_{\rm ej}}}

\slugcomment{To appear in The Astronomical Journal}

\shorttitle{A New Paradigm for Elliptical Galaxies}
\shortauthors{Trujillo et al.}

\begin{document}

\title{Evidence for a New Elliptical-Galaxy Paradigm: S\'ersic and Core Galaxies}

\author{I. Trujillo}
\affil{Max-Planck-Institut f\"ur Astronomie, K\"onigstuhl 17, D-69117 
Heidelberg, Germany}
\email{trujillo@mpia-hd.mpg.de}
\and
\author{{Peter Erwin}, {A. Asensio Ramos}}
\affil{Instituto de Astrof\'{\i}sica de Canarias, Calle V\'{\i}a L\'actea s/n, E-38200 La Laguna, Tenerife, Spain}
\email{erwin@ll.iac.es, aasensio@ll.iac.es}
\and
\author{Alister W. Graham}
\affil{Department of Astronomy, P.O. Box 112055, University of Florida, Gainesville, FL 32611}
\email{graham@astro.ufl.edu}

\begin{abstract} 

We fit the surface-brightness profiles of 21 elliptical galaxies using
both the S\'ersic function and a new empirical model which combines an
inner power law with an outer S\'ersic function.  The profiles are
combinations of deconvolved \textit{HST} profiles from the literature
and ellipse fits to the full WFPC2 mosaic images, and thus span a
radial range from $\sim 0\farcs02$ to $\sim$ twice the half-light
radius.  We are able to accurately fit the entire profiles using either
the S\'ersic function or our new model.  In doing so, we demonstrate
that most, if not all, so-called ``power-law'' galaxies are better
described as ``S\'ersic galaxies'' --- they are well modeled by the
three-parameter S\'ersic profile into the limits of \textit{HST}
resolution --- and that ``core'' galaxies are best understood as
consisting of an outer S\'ersic profile with an inner power-law cusp,
which is a downward deviation from the inward extrapolation of the
S\'ersic profile.  This definition of cores resolves ambiguities that
result when the popular ``Nuker law'' is fitted to the profiles of
ellipticals and bulges, particularly at lower luminosities.  We also
find that using the Nuker law to model core-galaxy nuclear profiles
systematically overestimates the core radii by factors of 1.5--4.5 and 
underestimates the inner power-law slope by $\sim 20$--40\% or more.

\end{abstract}


\section{Introduction} 
\label{sec:intro}

The availability of high-resolution imaging with the \textit{Hubble
Space Telescope} (\textit{HST}) has revolutionized the study of galaxy
centers.  Following up on early work by \citet{crane93},
\citet{kormendy94}, \citet{grillmair94}, \citet{jaffe94}, and
\citet{ferrarese94}, a series of papers by the ``Nuker team''
\citep{lauer95,byun96,gebhardt96,faber97} presented a detailed study of
the central regions of early-type galaxies (specifically, ellipticals
and the bulges of spiral galaxies).  They introduced a model for
fitting the radial surface-brightness profiles: a double power-law
with an adjustable transition region, dubbed the ``Nuker law'':
\begin{equation}
I(r) = I_b \, 2^{(\beta - \gamma)/\alpha}
\left(\frac{r}{r_b}\right)^{-\gamma} \left[1 + \left( \frac{r}{r_b}
\right)^{\alpha}\right]^{(\gamma - \beta)/\alpha}.
\end{equation}
The inner and outer power law exponents are $\gamma$ and $\beta$,
respectively; $I_{b}$ is the surface brightness at the core or
``break'' radius $r_b$, and $\alpha$ controls the sharpness of the
transition between the two power laws (larger $\alpha$ = sharper
transition).  They identified two distinct classes of galaxy centers:
``power-law'' galaxies, where the central surface brightness increases
into the limit of resolution with something like a steep power-law
profile; and ``core'' galaxies, where the luminosity profile turns over
at a fairly sharp ``break radius'' into a shallower power-law.
Ferrarese et al.\ and Faber et al.\ found evidence that global
parameters of early-type galaxies correlated with their nuclear
profiles: core galaxies tend to have high luminosities, boxy isophotes,
and pressure-supported kinematics, while power-law galaxies are
typically lower-luminosity and often have disky isophotes and
rotationally supported kinematics.

The Nuker-law parameterization of galaxy centers has subsequently
enjoyed a great deal of popularity, including extensive studies using
WFPC2 and NICMOS
\citep[e.g.,][]{rest01,quillen00,ravindranath01,laine03}, and
extensions to early- and late-type spirals
\citep[e.g.,][]{carollo98,seigar02}.  These more recent studies have,
however, suggested that the clear core/power-law dichotomy found by the
Nuker team may not be so clear after all.  In addition, almost all the
studies using \textit{HST} data and Nuker-law fits have left unanswered
a key question: how does the nuclear part of a bulge or elliptical,
seemingly well fit by a double power-law, connect to the outer
profiles of such systems, which are generally well fit by the
\citet{sersic} $r^{1/n}$ function?  In our first paper
\nocite{paper1}(Graham et al.\ 2003a, hereafter Paper~I), we discussed
some of the systematic problems and ambiguities which can arise when
using a double power-law model to fit galaxy light profiles, and
suggested a new hypothesis and a new model which might resolve some of
these problems.  The hypothesis has two parts: first, that the nuclear
(\textit{HST}-resolved) profiles of most lower-luminosity hot systems,
including the power-law galaxies, are simply inward extensions of each
galaxy's outer profile, best modeled with a S\'ersic function; second,
that core galaxies are best modeled with our new function, an outer
S\'ersic function with a break to an inner power-law.  In this paper,
we make an empirical test of this proposed solution, by modeling the
\textit{entire} light profiles of a sample of elliptical galaxies.

In what follows, we first review some of the problems stemming from
the use of the Nuker law, including the problem of how best to
identify genuine cores in galaxies (Section~\ref{sec:nuker-problems}); 
readers familiar with these issues can probably skip this section. 
We then discuss our sample selection, data reduction and analysis, and
the source of the profiles used (Section~\ref{sec:data}).  In
Section~\ref{sec:models}, we discuss the S\'ersic model and our new
model for core-galaxy profiles.  Section~\ref{sec:fits} presents
criteria for identifying core galaxies, and for discriminating between
core and S\'ersic profiles.  We also present the results of our fits
to the galaxy profiles and compare their fidelity to the profiles with
that of the Nuker-law fits.  Some of the implications are discussed in
Section~\ref{sec:discuss}, and we conclude with a brief summary in
Section~\ref{sec:summary}.  Finally, several useful mathematical
expressions related to our new model are presented in the Appendix.

\section{Some Outstanding Issues} 
\label{sec:nuker-problems}

\subsection{Relating Nuclear Surface Brightness Profiles to Outer 
Profiles}

The progress engendered by the use of \textit{HST} data and the Nuker
law has tended to encourage a disconnect between the inner and outer
regions of galaxies, which are studied separately and parameterized in
different fashions.  This is in part due to the fact that early
\textit{HST} studies using the first-generation Planetary Camera
generally provided useful data only for $r \lesssim 10\arcsec$
\citep[e.g.][]{lauer95}, so that only the nuclear region could be
studied.  But it is also due to the fact that the Nuker law does
\textit{not} describe the light profiles outside this region well, even
for ``single-component'' galaxies like ellipticals 
\citep[e.g.,][]{byun96}.

Meanwhile, there has been significant progress in understanding the
luminosity structure \textit{outside} the nuclear regions.  These
``global'' surface brightness profiles are usually well described with
\nocite{sersic}S\'ersic's (1968) $r^{1/n}$ law, a generalization of
\nocite{deV59}de Vaucouleurs' (1959) $r^{1/4}$ law.  This has been
shown to be true for both luminous ellipticals
\citep[e.g.,][]{capaccioli87,caon93,graham96} and dwarf ellipticals
\citep[e.g.,][]{davies88,cellone94,young94,durrell97,binggeli98,graham-guzman03}
and for the bulges of disk galaxies
\citep{andredakis95,seigar98,khosroshahi00b,graham01-disks,balcells03,macarthur03}.
There is now good evidence that the \textit{shape} of the overall
surface-brightness profile, as parameterized by the S\'ersic index $n$,
correlates with numerous (model-independent) elliptical and bulge
properties: the total luminosity, the central surface brightness, the
effective radius, and the central velocity dispersion
\citep{graham01-tgc,mollenhoff01,graham02-mn}.  It also correlates
extremely well with the mass of central supermassive black holes
\citep{graham01-smbh,erwin03-carnegie}.  This clearly points to
connections between the global distribution of stars in ellipticals and
bulges and the properties of their nuclear regions, and makes it more
important than ever to understand how the nuclear regions connect to
the outer parts of galaxies.

\subsection{The Ambiguity of Current Core and Power-law Definitions}

A second problem is the ambiguity of ``core'' versus ``power-law''
definitions, and the apparent unraveling of the clear distinction
between (high-luminosity) core and (lower-luminosity) power-law
galaxies reported by \citet{faber97}.  \citet{rest01} and
\citet{ravindranath01} have found several examples of ``intermediate''
galaxies ($0.3 < \gamma < 0.5$; see, e.g., Fig.~3 of Ravindranath et
al.); it is not clear where these galaxies fit into the core/power-law
scheme.  Taking a slightly different tack, \citet{carollo97} argued for
a general trend of $\gamma$ versus absolute magnitude for ellipticals,
with more luminous galaxies having shallower slopes: this roughly
matches the trend found by \nocite{faber97}Faber et al.\ 1997, but
without splitting the galaxies into core and power-law categories.
However, subsequent investigation of lower-luminosity systems,
particularly bulges in late-type galaxies and dwarf ellipticals, has
shown a \textit{reversal} of this trend: for low-luminosity systems,
luminosity and inner power-law slope are \textit{anti}-correlated
\citep[][ especially their Fig.~4]{stiavelli01}.  This has also been
portrayed as a dichotomy between more luminous ``$R^{1/4}$'' bulges,
with high $\gamma$, and less luminous ``exponential'' bulges, which
tend to have $\gamma < 0.3$ \citep{seigar02}.

\begin{figure}
\begin{center}
\includegraphics[scale=0.5]{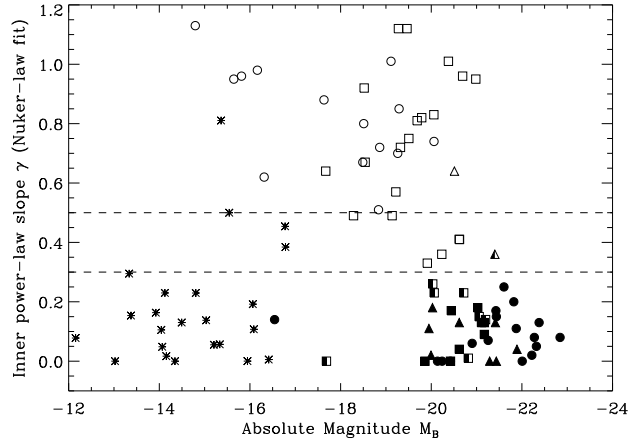}
\end{center}
\vfill \caption{The problem of how to identify ``cores'': inner
logarithmic slope $\gamma$, from Nuker-law fits to \textit{HST}
profiles, versus absolute magnitude $M_{B}$ for dwarf ellipticals from
Stiavelli et al.\ (2001, asterisks) and regular ellipticals from Faber
et al.\ (1997, circles), Rest et al.\ (2001, boxes), and Ravindranath
et al.\ (2001, triangles).  Filled symbols are core galaxies and
half-filled symbols are ``intermediate'' galaxies, according to the
authors of each study; Stiavelli et al.\ do not make core/non-core
classifications.  Total $B$ magnitudes are from LEDA, distances are
from Tonry et al.\ (2001) or LEDA (corrected for Virgo infall and
assuming $H_{0} = 75$ \kmskpc); for Virgo cluster galaxies without
measured distances, we assume $D = 15.3$ Mpc \citep{freedman01}.  Only
galaxies with Nuker-fit break radii $r_{b} \geq 0\farcs16$ are
plotted, so all galaxies with $\gamma < 0.3$ (lower dashed line) are
``core'' galaxies according to the standard definition
\citep{lauer95,faber97}; galaxies with $\gamma > 0.5$ (upper dashed
line) are ``power-law'' galaxies in the same scheme.}
\label{fig:gamma-core}

\end{figure}

To dramatize this problem, we plot $\gamma$ versus $M_{B}$ in
Figure~\ref{fig:gamma-core} for ellipticals spanning a wide range of
absolute magnitudes, from the brightest core galaxies of Faber et al.\
(1997) down to the faint dwarf ellipticals of Stiavelli et al.\ (2001);
a similar figure can be found in \citet{graham-guzman03}.  We indicate
the boundaries for core and power-law galaxies, according to
\citet{faber97}; all galaxies plotted have well-resolved ``cores''
($r_{b} \ge 0\farcs16$).  Two things stand out: first, there are
numerous ``intermediate'' objects, so that the rather clear distinction
reported by Faber et al.\ --- that systems with small $\gamma$ are high
luminosity, while systems with large $\gamma$ are lower luminosity ---
has become murky.  Second, if we apply the standard definition of a
core, then fully 21 of the 25 dwarf ellipticals of \citet{stiavelli01}
have cores!  Similarly, 12 of 38 spiral bulges (not plotted) studied in
the optical by \citet{carollo98} and 10 of 45 bulges studied in the
near-IR by \citet{seigar02} meet the standard criteria for having
cores.\footnote{Note that these authors do not classify centers into
core/power-law categories, and so do not actually label these ``core''
galaxies.} Either both low- and high-luminosity galaxies --- but
\textit{not} intermediate-luminosity systems --- have cores, or we need
a less problematic way of identifying cores.

As we showed in \nocite{paper1}Paper~I, this kind of ambiguity arises
automatically if the surface-brightness profile is exponential or
nearly so (i.e., a S\'ersic function with $n \lesssim 2$): when plotted
in log-log space --- and when fit with a double power-law such as the
Nuker law --- such profiles will seem to have cores.  Since dwarf
ellipticals and the bulges of many galaxies have profiles which are
well fit by S\'ersic functions with small $n$ (see references above),
this is clearly a concern.  The argument that S\'ersic profiles only
apply to the \textit{outer} parts of profiles (that is, outside the
region typically imaged by \textit{HST}) is not tenable.
\citet{geha02} and \citet{graham-guzman03} were able to fit the
\textit{HST} profiles of dwarf ellipticals using S\'ersic profiles
(plus optional nuclear components).  In addition, \citet{jerjen00}
found that the fully resolved surface-brightness profiles of Local
Group dwarf spheroidals --- which they show to be primarily the
low-luminosity extension of the dwarf ellipticals --- are quite well
fit by S\'ersic profiles \citep[see also][]{caldwell99}.

\begin{figure}
\begin{center}
\includegraphics[scale=0.45]{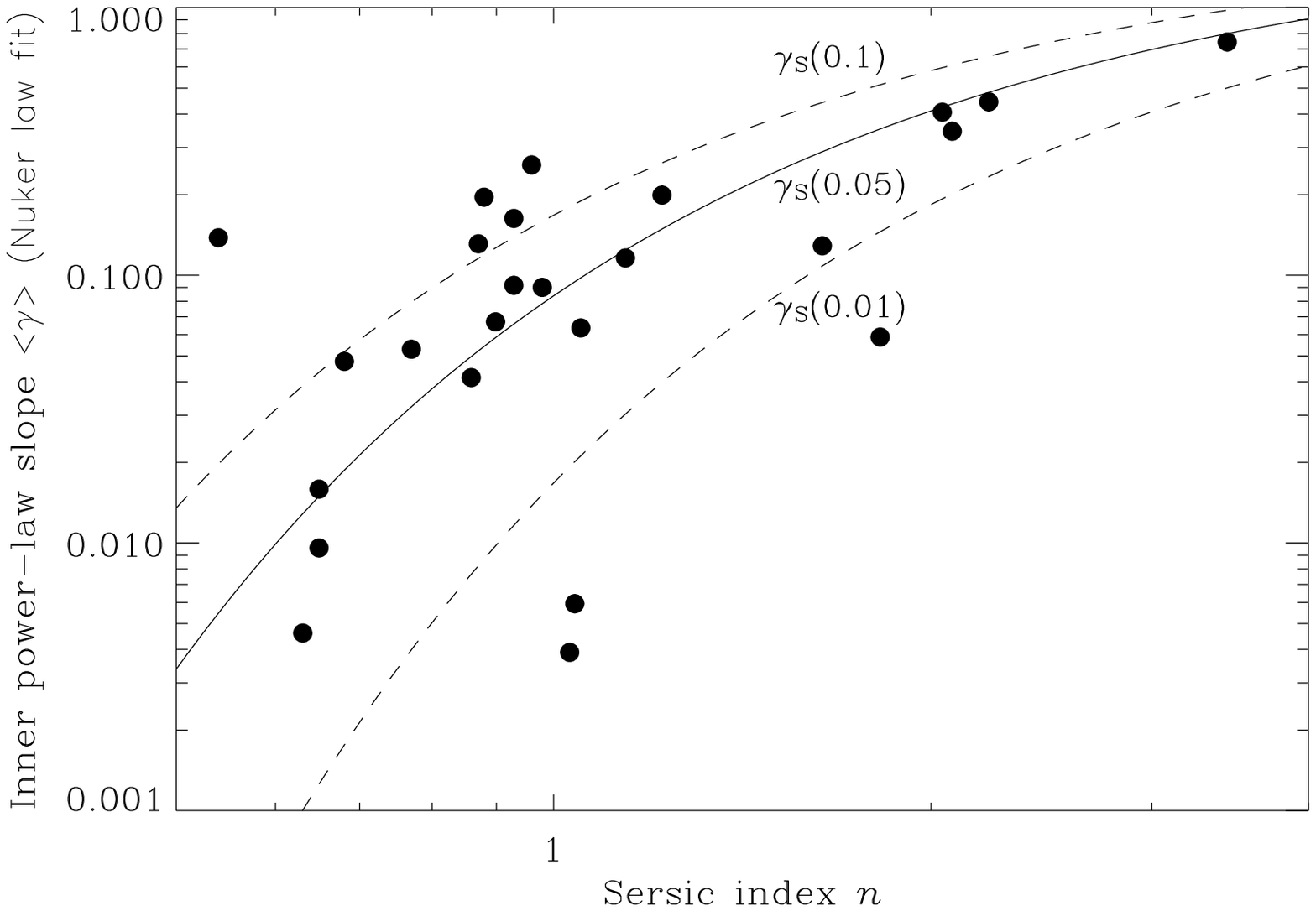}
\end{center}
%
\caption{Inner logarithmic slope $<\gamma>$ (from Nuker-law fits,
averaged over $r = 0\farcs1$--0\farcs5) versus the S\'ersic index $n$
for the dwarf ellipticals of \citet{stiavelli01}.  Also plotted are
curves showing the logarithmic slope of the S\'ersic function at
different fractions of the half-light radius (0.01, 0.05, and 0.1
$r_{e}$), derived using Eq.~\ref{sersicgamma}.}
\label{fig:de-gamma}

\end{figure}

Since the S\'ersic shape parameter $n$ is correlated with luminosity
\citep[e.g.,][]{caon93,jerjen00,graham-guzman03} and with
central velocity dispersion \citep{graham01-tgc,graham02-mn}, we have
a natural explanation for the correlation between $\gamma$ and
luminosity: S\'ersic profiles observed from the ground
continue inward into the regions resolved by \textit{HST}, so that
galaxies with larger $n$ (higher luminosities) will
have larger\footnote{Eq.~\ref{sersicgamma} shows the relation between
$\gamma$ and $n$ for a S\'ersic profile.} $\gamma$.
Figure~\ref{fig:de-gamma} shows that this is supported by the
S\'ersic fits and $\gamma$ measurements of \citet{stiavelli01}: dwarf
ellipticals with larger values of $n$ have larger values of $\gamma$,
in line with what we expect from S\'ersic profiles observed at small
radii.  In Section~\ref{sec:fits} we show that the inner regions of
higher-luminosity, power-law ellipticals (high $\gamma$) are well fit
by S\'ersic functions with large $n$ which simultaneously fit the
outer profiles.

But where does that leave core galaxies?  The results of
\citet{gebhardt96} and \citet{faber97} strongly suggest that the
low-$\gamma$ cores identified in these galaxies are genuine, physically
distinct structures; indeed, some of these cores were well-known from
high-resolution, ground-based imaging
\nocite{kormendy85,lauer85,lauer95}(e.g., Kormendy 1985; Lauer 1985;
see the discussion in Lauer et al.\ 1995).  The outer profiles of
\textit{high}-luminosity ellipticals, those most likely to have such
cores, have large values of $n$, so the inner slope $\gamma$ should be
large, the opposite of what is observed.  This means that cores in
bright ellipticals are clear deviations from the outer (S\'ersic)
profiles, and suggests a more natural way of identifying cores: a
downward \textit{deviation}, with shallow logarithmic slope, from a
galaxy's outer S\'ersic profile.  This would resolve the ambiguity we
noted above: illusory ``cores'' in low-luminosity systems (produced by
fitting a double-power law to low-$n$ S\'ersic profiles) cannot be
confused with \textit{true} cores in high-luminosity systems.  In
Section~\ref{sec:fits}, we show that this is indeed a viable approach:
the complete profiles of high-luminosity \textit{core} galaxies are
\textit{not} well fit by a single S\'ersic profile, but \textit{are}
well fit by our new model, which joins a single, inner power-law
profile to an outer S\'ersic profile.

\section{Sample Selection, Data Reduction, and Generation of Profiles} 
\label{sec:data}

\subsection{Sample Selection} 
\label{sec:sample}

For this study, we needed a set of galaxies with \textit{HST}
observations of their central regions, as well as observations of the
outer parts of the galaxies.  Ideally, we want to compare our results
with those of previous studies which used Nuker-law fits to analyze and
classify the galaxies.  This drove us to concentrate on the two largest
\textit{HST} studies of early-type galaxies: the WF/PC1 study of the
Nuker team \citep{lauer95,byun96}, and \citet{rest01}, which used
WFPC2.  In both cases, the authors presented deconvolved profiles
derived from ellipse fits to the Planetary Camera chips; \citet{rest01}
also present values at very small radii derived directly from
individual pixel values.  Since these are the data which the Nuker team
and Rest et al.\ use for their Nuker-law fits and classifications, it
made sense for us to use them as well.

The problem then became finding suitable profiles for the galaxies
outside the region imaged by the PC chips ($r \gtrsim 20\arcsec$).  To
minimize problems which might arise from combining profiles from
different filters, we needed $V$-band images to go with the F555W
profiles from \citet{lauer95} and $R$-band images to go with the F702W
profiles from \citet{rest01}.  We also wanted images with fairly high
resolution, to avoid any possible changes in curvature induced by
trying to match ground-based profiles with poor seeing to the
high-resolution \textit{HST} images.  The simplest solution to both of
these requirements was to use \textit{HST} images --- in particular,
WFPC2 images obtained using the same filters.  Although the WFPC2 array
is missing almost a quarter of its field, the overall field of view is
$\approx 2.6\arcmin$, which is sufficient to cover smaller galaxies;
for larger galaxies, we can still sample most of the profile with the
ellipse fits.  In addition, the very low background in \textit{HST}
images means that we are less vulnerable to sky subtraction errors,
which can affect the outer profiles.  In practice, we found the
following restrictions worked best: major axis $< 4\arcmin$ and minor
axis $< 3\arcmin$, using the $\mu_{B} = 25$ dimensions from
\nocite{rc3}de Vaucouleurs et al.\ (1991, hereafter RC3).

The decision to use \textit{HST} images makes the match with the inner
profiles of \citet{rest01} particularly good: it means that we are
using the exact same F702W images they used.  For the F555W profiles
of the Nuker team, we searched the \textit{HST} archive for WFPC2
images in the same filter (the F555W filters of the two cameras are
not precisely identical, but the differences are too small to matter). 
There were somewhat fewer of these, so most of the galaxies we analyze
are from the Rest et al.\ sample.

Finally, we decided to examine only elliptical galaxies.  Although the
bulges of disk galaxies are known to be well fit by the S\'ersic model,
extracting the actual bulge profile means making bulge-disk
decompositions.  While not a significant problem for some galaxies, it
does add some uncertainty, since we could end up fitting a
one-dimensional profile with as many as \textit{eight} free parameters
(disk scale length and central surface brightness + five or six
parameters for our new model).  In the future, we do plan to analyze
the bulges of disk galaxies using our new model, but for the purposes
of this study we wanted to simplify matters and eliminate as much
ambiguity as possible.

Thus, we selected only elliptical galaxies from the samples of the
Nuker team and \citet{rest01}.  This meant not just selecting those
galaxies classified as elliptical, but also ensuring that they were,
in fact, true ellipticals with no significant disk component.  A
number of nominal E galaxies showed signs of having significant outer
disks, suggesting that they may well be misclassified E/S0 or S0
galaxies.  Our criteria included kinematic evidence from the
literature, ellipse fits to the WFPC2 mosaic images, bulge+disk
decompositions using the extra-nuclear ($r > 1 \arcsec$) part of the
profiles, and the presence of substructures such as rings and bars,
which are evidence for disks massive enough to be self-gravitating. 
Appendix~\ref{app:rejected} discusses rejected galaxies on a
case-by-case basis.  The remaining 21 galaxies, which we judged to be
bona-fide ellipticals, are listed in Table~\ref{tab:sample}.

The angular size limits and the nature of the previous samples we draw
on mean that the galaxies in Table~\ref{tab:sample} span a limited
range in absolute magnitude.  Happily, this narrow magnitude range
ends up bracketing the overlap between core and power-law galaxies,
and we have roughly equal numbers of each.

\begin{deluxetable}{cccccccccc}
\tabletypesize{\scriptsize}
\tablecolumns{10}
\tablewidth{0pc}
\tablecaption{The galaxy sample and global parameters\label{tab:sample}}
\tablehead{
Galaxy   & Type & $B_{T}$ & $M_B$& Distance & source & $V_{\rm vir}$ &Innermost Data & $\sigma$ & Profile Type\\
   (1)    & (2)  &  (3) & (4)  & (5) & (6) & (7)  & (8) & (9) & (10)}
\startdata
NGC 1426 & E4   &  12.62 &  -19.29 &  24.1 & 3 &	1232 & $0\farcs25$ / 29.2 pc & 155 & $\setminus$\\
NGC 1700 & E4   &  11.87 &  -21.36 &  44.3 & 3 &	3800 & 0.09 / 19.3 & 243 & $\setminus$\\
NGC 4458 & E0-1 &  12.86 &  -18.32 &  17.2 & 3 &	 768 & 0.13 / 10.8 & 106 & $\setminus$\\
NGC 5845 & E:   &  13.24 &  -18.83 &  25.9 & 3 &	1634 & 0.02 /  2.8 & 244 & $\setminus$ \\
\cutinhead{From Rest et al.\ (2001)}\\
NGC 2634 & E1:  &  12.93 &  -19.69 &  33.4 & 3 &	2539 & 0.10 / 16.2 & 172 & $\setminus$ \\
NGC 2872 & E2   &  12.67 &  -20.44 &  41.9 & 4 &	3143 & 0.49 / 99.5 & 284 & $\setminus$ \\
NGC 2986 & E2   &  11.41 &  -20.89 &  28.9 & 4 &	2170 & 0.02 /  2.8 & 260 & $\cap$\\
NGC 3078 & E2-3 &  11.75 &  -20.98 &  35.2 & 3 &	2339 & 0.63 /  108 & 237 & $\setminus$\\
NGC 3348 & E0   &  11.71 &  -21.36 &  41.2 & 4 &	3092 & 0.02 /  4.0 & 239 & $\cap$\\
NGC 3613 & E6   &  11.70 &  -20.62 &  29.1 & 3 &	2246 & 0.05 /  7.1 & 205 & $\cap$\\
NGC 4168 & E2   &  12.00 &  -20.45 &  30.9 & 3 &	2396 & 0.12 / 18.0 & 186 & $\cap$\\
NGC 4291 & E3   &  12.23 &  -19.40 &  26.2 & 3 &	2047 & 0.04 /  5.1 & 278 & $\cap$\\
NGC 4478 & E2   &  12.07 &  -19.22 &  18.1 & 3 &	1485 & 0.02 /  1.8 & 143 & $\setminus$\\
NGC 5017 & E+?  &  13.18 &  -19.43 &  33.3 & 4 &	2495 & 0.33 / 53.3 & 174 & $\setminus$\\
NGC 5077 & E3-4 &  12.12 &  -20.70 &  36.7 & 4 &	2752 & 0.14 / 24.9 & 273 & )\\
NGC 5557 & E1   &  11.96 &  -21.34 &  45.7 & 4 &	3427 & 0.02 /  4.4 & 259 & )\\
NGC 5576 & E3   &  11.80 &  -20.23 &  25.5 & 3 &	1565 & 0.02 /  2.5 & 190 & $\setminus$\\
NGC 5796 & E0-1 &  12.36 &  -20.61 &  39.3 & 4 &	2950 & 0.02 /  3.8 & 290 & $\setminus$\\
NGC 5831 & E3   &  12.62 &  -19.55 &  27.2 & 3 &	1740 & 0.02 /  2.6 & 168 & $\setminus$\\
NGC 5903 & E2   &  11.48 &  -21.17 &  33.9 & 3 &	2466 & 0.02 /  3.3 & 217 & $\cap$\\
NGC 5982 & E3   &  11.88 &  -21.25 &  42.2 & 4 &	3168 & 0.02 /  4.1 & 256 & $\cap$\\
\enddata

\tablecomments{Global parameters for the galaxies in our sample:
Col.~(1): Galaxy name.  Col.~(2): Morphological type from RC3. 
Col.~(3): Total apparent $B$-band magnitude, corrected for Galactic
extinction and redshift, from LEDA \citep[see][]{paturel97}. 
Col.~(4): Absolute $B$-band magnitude, using distance from column~5. 
Col.~(5): Distance in Mpc.  Col.~(6): Sources for the distances: 3 =
SBF distance from \citet{tonry01}, 4 = corrected radial velocity
(col.~[7]) and $H_{0} = 75$ km s$^{-1}$ kpc$^{-1}$.  Col.~(7): Radial
velocity (in km s$^{-1}$), corrected for Virgocentric infall, from
LEDA \citep[infall model in][]{paturel97}.  Col.~(8): Radius of
innermost data point used in fits, in arc seconds and in parsecs. 
Col.~(9): Central velocity dispersion in km s$^{-1}$, from \citet{mcelroy95}.  Col.~(10): Original \textit{HST} inner profile classification from
Nuker-law fits, from either \citet{lauer95} or \citet{rest01};
$\setminus$, ), and $\cap$ indicate power--law, intermediate, and core
galaxies, respectively.}

\end{deluxetable}

\subsection{Data Reduction and Profile Matching} 
\label{sec:reduction}

The WFPC2 images were retrieved from the \textit{HST} archive with
standard on-the-fly calibration.  Multiple exposures were combined
using the \texttt{crrej} task within \textsc{iraf}.  Alignment of
different exposures was checked using coordinates of bright stars and
galaxy nuclei; if the offset was $\lesssim 0.2$ pixels in the PC chip,
then the images were combined without shifting.  (Since we use the
published profiles of \nocite{lauer95}Lauer et al.\ 1995 and
\nocite{rest01}Rest et al.\ 2001 for $r \lesssim 10\arcsec$, we do not
need highly accurate alignment.)  We then made mosaic images from the
combined exposures using the \texttt{wmosaic} task.  Sky subtraction
was based on the average of median values from several $10 \times 10$
pixel boxes, located well away from the galaxy.  In some cases, there
was evidence that galaxy light was present even at the edges of the WF
chips, so the outermost one or two points in the profiles may not be
very reliable (the effect appears to be significant only for NGC~2986).

We derived surface-brightness profiles from the sky-subtracted mosaic
images by fitting ellipses to the isophotes with the \textsc{iraf} task
\texttt{ellipse}, using logarithmic spacing and median filtering.  The
mosaic images were first masked to exclude the missing quadrant and the
gaps between the individual chips, as well any bright foreground stars
or other galaxies.  Results of ellipse fits are shown in
Appendix~\ref{app:figs}.

The resulting major-axis profiles were then combined with the
published, deconvolved major-axis profiles for the inner regions, from
\citet{lauer95} and \citet{rest01}; these inner profiles typically
extend to semi-major axis $a \approx 10$--20\arcsec.  We matched our
outer, mosaic-based profiles to these inner profiles using the overlap
at $2\arcsec \leq a \leq 10\arcsec$.  This is sufficiently outside the
center that differences due to resolution effects are minimized.  The
profile from the mosaic was only used for radii outside the literature
profiles, except for some of the profiles from Rest et al., where we
added points from the mosaic profile to fill in gaps in their profiles
at $r > 2\arcsec$, to create composite profiles that were more evenly
spaced in logarithmic radius.  For the Rest et al.\ profiles, we
attempted to sample the inner part of their profiles with approximately
the same spacing as our mosaic profiles, again with the aim of
producing composite profiles that are more-or-less evenly spaced in
logarithmic radius.  If the original studies excluded values at small
radii from the fits (as indicated by Figure~3 of Byun et al.\ 1996 and
Figure~8 of Rest et al.)  --- due to the presence of distinct nuclei or
strong dust absorption --- then we also excluded those
points.\footnote{The exception is NGC~5845, where we were only able to
reproduce the original Nuker-law fit (and rms residuals) of Byun et
al.\ by including \textit{all} of the inner points.} The combined
profiles can be seen in Section~\ref{sec:fits}.

\section{Models for Galaxy Light Profiles} 
\label{sec:models}

The \citet{sersic} model can be defined as: 
\begin{equation} 
I(r) = I(0) \exp[-b_{n}(r/r_e)^{1/n}],
\end{equation}
with $I(0)$ being the central intensity, $r_e$ the scale radius (=
half-light radius), and $n$ the shape parameter controlling the
overall curvature; when $n = 1$, this reduces to an exponential, while
$n = 4$ gives the traditional \citet{deV59} $r^{1/4}$ profile.  The
quantity $b_{n}$ is a function of the shape parameter $n$, chosen to
ensure that the scale radius encloses half of the total luminosity. 
The evaluation of $b_{n}$ can be found in Eqn.~\ref{eq:sersicb}.

Our new model, introduced in \nocite{paper1}Paper~I and
\citet{carnegie}, is analogous to the Nuker law, but uses the S\'ersic
model for the outer part of the profile (see Paper~I for some
representative plots).  This model, which we will refer to as
``\bomba{},'' is
\begin{equation}
I(r) = I^{\prime}
\bigg[1 + \bigg(\frac{r_b}{r}\bigg)^{\alpha}\bigg]^{\gamma/\alpha}
\exp\bigg[-b \bigg(\frac{r^\alpha+r_b^\alpha}{r_e^\alpha}\bigg)^{1/(n\alpha)}
\bigg],
\label{alphafree}
\end{equation}
with 
\begin{equation}
I^{\prime} = I_b \; 2^{-\gamma/\alpha} \exp[\, b \, 2^{1/\alpha n} \,
(r_b/r_e)^{1/n}].
\end{equation}
The parameters have the same general meaning as in the S\'ersic or
Nuker laws: the break radius $r_b$ is the point at which the profile
changes from one regime to another, $\gamma$ is the slope of the inner
power law region, $I_b$ is the intensity at the break radius, $\alpha$
controls the sharpness of the transition between the cusp and the
outer S\'ersic profile, $r_e$ is the effective radius of the profile,
and $n$ is the shape parameter of the outer S\'ersic part.  The
quantity $b$ is a function of the parameters $\alpha$, $r_b/r_e$,
$\gamma$, and $n$, and is defined in such a way that $r_e$ becomes the
radius enclosing half the light of the galaxy model (see
Appendix~\ref{app:math}).  If $\alpha \rightarrow \infty$, then the
transition from S\'ersic profile to power law at $r_{b}$ is infinitely
sharp, with no transition region.  In this limiting case, the model
can be written as:
\begin{equation}
I(r) = I_{b} \bigg[\bigg(\frac{r_b}{r}\bigg)^{\gamma}u(r_b-r) \, + \,
e^{b(r_b/r_e)^{1/n}}e^{-b(r/r_e)^{1/n}}u(r-r_b)\bigg],
\label{alpha-infty}
\end{equation}
with $u(x - a)$ being the Heaviside step function. 
Eq.~\ref{alpha-infty} can also be approximated using
Eq.~\ref{alphafree} with $\alpha \gtrsim $100.  \citet{carollo98}
introduced a more limited version of Eqn.~\ref{alphafree}, with a
non-adjustable transition region and an exponential instead of the
S\'ersic outer region.  (They used it to model --- generally without
success --- the profiles of low-luminosity, ``exponential'' bulges
with nuclear excesses, rather than those of the higher-luminosity
ellipticals which typically have cores.)

For the $\alpha \rightarrow \infty$ case, the relation between the
intensity at the effective radius $r_e$ and the intensity at the break
radius $r_b$, assuming that $r_{e} > r_{b}$, is given by:
\begin{equation}
I(r_e) = I_{b} \exp[b((r_b/r_e)^{1/n} - 1)]
\end{equation}
or, equivalently, 
\begin{equation}
\mu_e = \mu_b - 2.5 b ((r_b/r_e)^{1/n} - 1) \log e.
\end{equation}

The definition for $b$ in the general case ($\alpha =$ free) is
somewhat complex, though the necessary integrations can be done
numerically beforehand and interpolated for actual
fitting.\footnote{In the $\alpha = \infty$ case the definition of $b$
is simpler; see Eq.~A12.} A simpler, mathematically equivalent version
can be had if we replace $b$ by $b_{n}$ from the S\'ersic model, in
which case $r_{e} \rightarrow r_{es}$, the half-light radius of the
outer S\'ersic profile (i.e., considered as a complete S\'ersic
profile extending in to $r = 0$).\footnote{This is the version given
in Paper~I, where $r_{e}$ was used for what we term $r_{es}$ here.}
For unrealistically large cores (inner power-law regions), this
$r_{es}$ (and its corresponding $\mu_{es}$) will not be a good
approximation to the true $r_{e}$ and $\mu_{e}$ of the profile.  In
practice, as long as $r_{b} \ll r_{e}$ and $\alpha \gtrsim 1$, the
difference will almost certainly be much less than the uncertainty in
$r_{e}$ from the fitting process itself (see, e.g.,
\nocite{paper3}Paper~III).

The \bomba{} model in its general form has six free parameters, one
more than the Nuker law.  However, it is possible that when fitting
real galaxy profiles the parameter $\alpha$, which controls the
sharpness of the transition between outer S\'ersic and inner power-law
regimes, may not be necessary.  If a galaxy has a distinct (power-law)
core, then the transition to the outer S\'ersic profile could, in
principle, not be fully resolvable, and might be adequately modeled
using $\alpha = \infty$ (i.e., the sharp-transition model,
Eqn.~\ref{alpha-infty}).  The \textit{Nuker law} requires low values
of $\alpha$, for both core \textit{and} power-law galaxies, because
this is the only way to create the significant curvature needed to
reproduce the observed curvature of galaxy profiles.  But since the
S\'ersic part of our profile already models that curvature, we do not
automatically need a low-$\alpha$ transition.  There are additionally
two mathematical reasons for preferring the sharp-transition model. 
First, it reduces the number of free parameters in the model to five. 
Second, a smooth transition (low $\alpha$) distorts the meaning of the
other parameters, so that, for example, the logarithmic slope of the
inner profile is \textit{not} equal to $\gamma$ except at very small
radii (as discussed in Section~\ref{sec:nuker-problems}).

Thus, we use \textit{both} Eqns.~\ref{alphafree} and \ref{alpha-infty}
to model galaxy profiles.  Our hope, from the standpoint of simplicity
and more transparent meaning for the model parameters, is that the
sharp-transition model will be sufficient for core galaxies; as we
show in Section 5.3, this appears to be the case.

\section{Fits to Galaxy Profiles} 
\label{sec:fits}

\subsection{Fitting Techniques and Comparisons with Previous Fits} 

We fitted various models to the profiles using two standard nonlinear
least-squares techniques: the downhill simplex (``amoeba'') method,
and the Levenberg-Marquardt method \citep[see, e.g.,][]{nr}; many of
the profiles were also fit using a quasi-Newton algorithm
\citep{kahaner89}.  This went some way towards ensuring that our
results were not dependent on the peculiarities of a single method, or
its implementation.  In general, we found excellent agreement between
fits obtained with the three methods.  We also tried a variety of
starting parameters, to ensure that our fits did not get trapped in
local $\chi^{2}$ minima.  Following \citet{byun96}, we weighted all
points equally.

One test of our fitting methods is to see how well we reproduce the
original Nuker-law fits of \citet{byun96} and \citet{rest01}, if we
restrict the radial range to that of the published PC profiles.  In
general, we did fairly well at this.  There are minor differences
between our Nuker-law fits and those of Byun et al.\ (typically only
10--20\% in parameter values) because the latter performed their fits
to the equivalent radius ($r_{\rm eq} = \sqrt{a b}$) profiles, rather
than to the major-axis profiles as we do.  They also used the
(unpublished) cumulative $r \leq 0\farcs1$ flux as an additional
constraint on the fits in some cases.

We found similarly good agreement with the original Rest et al.\ fits
for about two-thirds of the galaxies drawn from their sample; but more
significant differences exist for the remainder.  There are two
probable reasons for this.  First, Rest et al.\ used a somewhat
complex scheme of weighting the data points by the errors, while we
weight all points equally.  Second, their deconvolved profiles are
often \textit{not} evenly sampled in logarithmic radius; this can have
the effect of giving more weight to points at smaller radii.  For
example, we get a much closer match to the their Nuker-law fit for
NGC~5576 if we fit to $a \leq 5\arcsec$ in our combined profile
instead of $a \leq 16\arcsec$, since there are few data points in
their deconvolved profile beyond $a = 5\arcsec$ (our combined profiles
have had any such gaps filled in with points from the ellipse fits to
the mosaic image, in order to produce more evenly sampled profiles). 
This dependence on the radial weighting is probably a manifestation of
the general radial sensitivity of Nuker-law fits
\nocite{paper1,paper3}(Papers~I and III), something supported by the
fact that when our fits differ significantly from those of Rest et
al., our $r_{b}$ values are always larger.

\subsection{Distinguishing Core from S\'ersic Profiles} 

The Nuker team devised a simple set of criteria for separating core
from power-law galaxies, based on fitting profiles with the Nuker law
\citep{lauer95,faber97}: if the Nuker-law break radius was large
enough to be well-resolved ($r_{b} \ge 0\farcs16$) and the inner
power-law slope was sufficiently flat ($\gamma \leq 0.3$), then the
galaxy was considered to have a core; otherwise, it was classed as
power-law \citep[or possibly as ``intermediate'';
e.g.,][]{rest01,ravindranath01,laine03}.

\begin{figure}
\begin{center}
\includegraphics[scale=0.88]{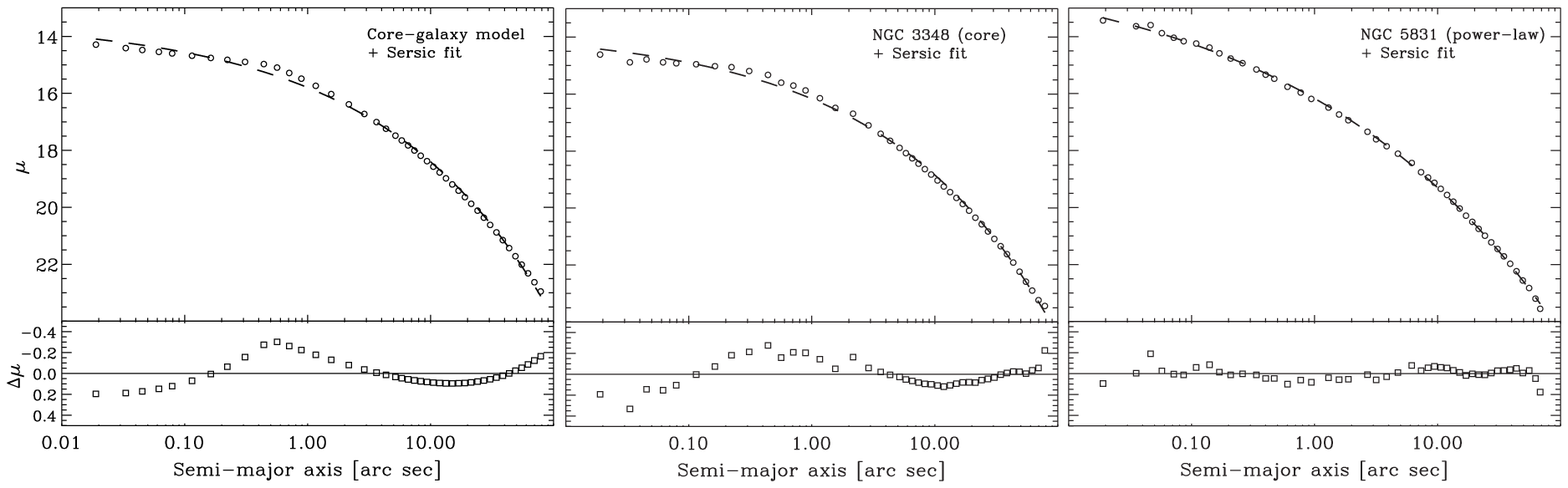}
\end{center}

\caption{How to identify core galaxies using the residuals of a
S\'ersic fit to the surface-brightness profile.  \textit{Left}: a
model profile for a core galaxy: a de Vaucouleurs profile ($r_{e} =
25\arcsec$) with a sharp break at $r_{b} = 0\farcs5$ to a power-law
core with $\gamma = 0.2$.  \textit{Middle}: profile of the core galaxy
NGC~3348.  \textit{Right}: profile of the power-law galaxy NGC~5831. 
For all three, we also show the best-fitting S\'ersic profile (dashed
line) and the residuals of the fit (boxes).  The characteristic
pattern of the residuals (compare model and NGC~3348 versus NGC~5831)
indicates a qualitative way of distinguishing core-galaxy
profiles.}\label{fig:core-residuals}

\end{figure}

Our approach is somewhat different: we want to determine when a galaxy
profile is best fit by one of two profiles, S\'ersic or \bomba, and
--- something which is in principle a separate issue --- whether the
galaxy has a core or not.  Which model provides a better fit can be
determined by comparing reduced $\chi^{2}$ values.  Galaxies which are
well fit with the S\'ersic profile do not, by our definition, have
cores.  However, just getting a significantly better fit with the
\bomba{} model does not necessarily indicate a \textit{core}.  For
example, a bright nuclear disk could add a distinct break to an
underlying S\'ersic profile; the composite would then be better fit by
the \bomba{} model, even though the overall elliptical/bulge profile
was still S\'ersic.  As suggested in \nocite{paper1}Paper~I,
\textit{we define a ``core'' as a downward deviation from the inward
extrapolation of the outer (S\'ersic) profile}.  Examples can be seen 
in Figures~\ref{fig:core-residuals} and \ref{fig:core-fits}.

After some experimentation, we settled on the following criteria for
clearly identifying core galaxies:
\begin{enumerate}
    \item Qualitative identification of cores: attempting to fit an
    idealized core galaxy with a S\'ersic profile produces a
    characteristic pattern in the residuals
    (Figure~\ref{fig:core-residuals}).  By fitting all galaxy profiles
    with the S\'ersic model and examining the residuals, we can
    \textit{qualitatively} identify core galaxies.
    
    \item Significantly better fit with \bomba{} (CS) than with
    S\'ersic models: \chinu(S\'ersic)$ \; > \; 2 \, \chinu($CS)
    indicates that the \bomba{} fit is clearly better, while
    \chinu(S\'ersic)$ \; \leq \; 1.2 \, \chinu($CS) indicates the
    S\'ersic profile is good enough.  Intermediate ratios are
    ambiguous cases, which we discuss further below.
    
    \item Potential cores must be both well-resolved and represented
    by enough data points.  Cases where the \bomba{} break radius is
    greater than the innermost data point are potentially
    non-S\'ersic profiles, but if the power-law regime is defined by
    only one or two data points, then its reality is dubious (and the
    inner slope $\gamma$ will be poorly defined).  Thus, for
    unambiguous core detection we require $r_{b} > r_{2}$, where
    $r_{2}$ is the second innermost data point in the profile.
    
    \item Finally, for a true core profile we require that the
    power-law slope be consistently $<$ the logarithmic slope of the
    S\'ersic fit inside break radius.\footnote{This applies to the 
    fitted data only; as $r \rightarrow 0$, the S\'ersic slope 
    $\rightarrow 0$ as well, but this happens well inside the 
    resolution limit for all our galaxies.}
\end{enumerate}

Non-core galaxies can then be divided into two classes: pure S\'ersic 
profiles, and problematic cases, the latter usually due to a 
significant extra component such as a bright nuclear disk.

Figures~\ref{fig:core-fits} and \ref{fig:sersic-fits} show the fits
for core and S\'ersic/ambiguous galaxies, respectively;
Table~\ref{tab:fits} lists the parameters of the fits.  The
classifications are based on \textit{our} fits, although, as we
discuss below, we reproduce the core/power-law classifications of
\citet{byun96} and \citet{rest01} almost perfectly.  For each galaxy
in the figures we show the best S\'ersic and \bomba{} fits to the
entire profile.  We also show the best Nuker-law fit \textit{to the
inner profile obtained from the PC chip}.  We do this because we wish
to compare how well a S\'ersic or \bomba{} fit to the \textit{entire}
profile manages to reproduce the \textit{inner} profile, where the
Nuker law has been used.  The relative goodness of the fits is given
in Table~\ref{tab:fits}, where we list the reduced chi-square values
\chinu{} for the S\'ersic and \bomba{} fits, and in
Table~\ref{tab:rms}, where we give rms residuals for all three types
of fit (S\'ersic, \bomba{}, and Nuker-law), evaluated in the inner
(PC) region.  Again, we do this so we can explicitly compare how well
the global S\'ersic or \bomba{} fit does at reproducing the inner
(\textit{HST}-resolved) part of the profile.

\begin{deluxetable}{ccccccccccc}
\tabletypesize{\scriptsize}
\tablecolumns{11}
\tablewidth{0pc}
\tablecaption{Structural Parameters\label{tab:fits}}
\tablehead{ Galaxy & $n$ &
$r_e$ & $I_e$ & $I_b$ & $r_b$ & $\gamma$ & $\alpha$ & \chinu
& NL-Fit Type & Notes\\
  (1)    & (2)  &  (3) & (4)  & (5)& (6) & (7) & (8) & (9) & (10) & 
  (11)}
\startdata
\cutinhead{Core Galaxies}\\
N2986  & 3.29 & 26.1  & 20.40 & ---   & ---  & ---  & ---      & 0.0353 &
$\cap$ & \\
       & 5.28 & 43.5  & ---   & 15.51 & 0.69 & 0.25 & 156.8    & 0.0108 &
& \\
       & 5.28 & 43.5  & ---   & 15.51 & 0.69 & 0.25 & $\infty$ & 0.0105 &
& \\
N3348  & 3.09 & 19.8  & 20.10 & ---   & ---  & ---  & ---      & 0.0172 &
$\cap$ & \\
       & 3.86 & 22.4  & ---   & 15.23 & 0.34 & 0.14 & 3.79     & 0.0017 &
& \\
       & 3.81 & 22.3  & ---   & 15.17 & 0.35 & 0.16 & $\infty$ & 0.0017 &
& \\
N4168  & 2.68 & 25.9  & 20.84 & ---   & ---  & ---  & ---      & 0.0058 &
$\cap$ & \\
       & 7.47 & 13.6  & ---   & 17.80 & 3.15 & 0.00 & 0.72     & 0.0012 &
& 1 \\
       & 3.12 & 29.2  & ---   & 16.66 & 0.72 & 0.22 & $\infty$ & 0.0016 &
& \\
N4291  & 3.75 & 15.7  & 19.88 & ---   & ---  & ---  & ---      & 0.0511 &
$\cap$ & \\
       & 5.58 & 18.2  & ---   & 14.56 & 0.36 & 0.11 & 4.42     & 0.0072 &
& \\
       & 5.44 & 18.1  & ---   & 14.48 & 0.37 & 0.14 & $\infty$ & 0.0073 &
& \\
N5557  & 3.74 & 23.3  & 20.47 & ---   & ---  & ---  & ---      & 0.0119 &
) & \\
       & 4.63 & 27.6  & ---   & 14.69 & 0.17 & 0.09 & 1.61     & 0.0019 &
& \\
       & 4.37 & 26.8  & ---   & 14.74 & 0.23 & 0.23 & $\infty$ & 0.0024 &
& \\
N5903  & 2.96 & 31.2  & 20.94 & ---   & ---  & ---  & ---      & 0.0346 &
$\cap$ & \\
       & 5.39 & 57.5  & ---   & 16.30 & 0.84 & 0.11 & 3.11     & 0.0058 &
& \\
       & 5.09 & 54.2  & ---   & 16.20 & 0.86 & 0.15 & $\infty$ & 0.0063 &
& \\
N5982  & 3.24 & 20.5  & 20.04 & ---   & ---  & ---  & ---      & 0.0210 &
$\cap$ & \\
       & 4.20 & 24.4  & ---   & 14.86 & 0.25 & 0.05 & 2.50     & 0.0012 &
& \\
       & 4.06 & 24.0  & ---   & 14.81 & 0.28 & 0.11 & $\infty$ & 0.0016 &
& \\

\cutinhead{Possible Core Galaxies}
N3613  & 3.63 & 34.2  & 20.63 & ---   & ---  & ---  & ---      & 0.0124 &
$\cap$ & \\
       & 3.89 & 37.1  & ---   & 14.70 & 0.13 & 0.00 & 4.61     & 0.0093 &
& \\
       & 3.87 & 36.9  & ---   & 14.71 & 0.15 & 0.09 & $\infty$ & 0.0092 &
& \\
N5077  & 3.56 & 21.7  & 20.34 & ---   & ---  & ---  & ---      & 0.0453 &
) & \\
       & 3.84 & 22.4  & ---   & 15.03 & 0.22 & 0.00 & 2.37     & 0.0288 &
& \\
       & 3.78 & 22.3  & ---   & 15.21 & 0.36 & 0.29 & $\infty$ & 0.0285 &
& \\

\cutinhead{S\'ersic Galaxies}\\

N1426  & 4.95 & 35.5  & 22.15 & ---   & ---  & ---  & ---      & 0.0014 &
$\setminus$ & 2 \\
       & 5.33 & 38.0  & ---   & 16.97 & 1.21 & 0.81 & 12.2     & 0.0011 &
& \\
       & 5.27 & 37.7  & ---   & 16.88 & 1.11 & 0.81 & $\infty$ & 0.0011 &
& \\
N1700  & 5.99 & 34.4  & 21.95 & ---   & ---  & ---  & ---      & 0.0039 &
$\setminus$ & \\
       & 5.98 & 34.4  & ---   & 13.19 & (0.03) & 0.00 & 33.9     & 0.0042 &
& \\
       & 5.98 & 34.4  & ---   & 12.94 & (0.02) & 0.00 & $\infty$ & 0.0042 &
& \\
N2634  & 4.54 & 18.1  & 21.10 & ---   & ---  & ---  & ---      & 0.0050 &
$\setminus$ & \\
       & 5.01 & 18.5  & ---   & 17.05 & 1.51 & 0.85 & 8.90     & 0.0038 &
& \\
       & 4.99 & 18.5  & ---   & 17.04 & 1.52 & 0.86 & $\infty$ & 0.0038 &
& \\
N2872  & 4.56 & 21.1  & 20.94 & ---   & ---  & ---  & ---      & 0.0048 &
$\setminus$ & \\
       & 4.56 & 21.1  & ---   & 11.54 & (0.00) & 0.28 & 17.2     & 0.0052 &
& \\
       & 4.56 & 21.1  & ---   & 11.39 & (0.00) & 0.30 & $\infty$ & 0.0051 &
& \\
N3078  & 4.37 & 22.9  & 20.36 & ---   & ---  & ---  & ---      & 0.0017 &
$\setminus$ & \\
       & 4.37 & 22.9  & ---   & 13.80 & (0.09) & 0.29 & 8.33     & 0.0018 &
& \\
       & 4.37 & 22.9  & ---   & 13.90 & (0.10) & 0.27 & $\infty$ & 0.0018 &
& \\
N4458* & 10.1 & 49.0  & 24.06 & ---   & ---  & ---  & ---      & 0.0373 &
$\setminus$ & 3 \\
       & 10.1 & 49.1  & ---   & 13.65 & (0.06) & 0.02 & 15.39    & 0.0403 &
& \\
       & 10.1 & 49.1  & ---   & 14.18 & (0.10) & 0.30 & $\infty$ & 0.0392 &
& \\
N4478* & 3.11 & 12.9  & 19.43 & ---   & ---  & ---  & ---      & 0.0484 &
$\setminus$ & 3 \\
       & 3.11 & 12.9  & ---   & 13.03 & (0.00) & 0.65 & 38.00    & 0.0521 &
& \\
       & 2.30 & 12.5  & ---   & 16.29 & 1.30 & 0.69 & $\infty$ & 0.0202 &
& \\
N5017  & 5.11 & 11.8  & 20.44 & ---   & ---  & ---  & ---      & 0.0082 &
$\setminus$ & \\
       & 5.11 & 11.8  & ---   & 13.92 & (0.10) & 0.37 & 34.5     & 0.0092 &
& \\
       & 5.11 & 11.8  & ---   & 10.00 & (0.00) & 0.62 & $\infty$ & 0.0092 &
& \\
N5576  & 4.74 & 32.0  & 20.63 & ---   & ---  & ---  & ---      & 0.0084 &
$\setminus$ & \\
       & 4.89 & 33.7  & ---   & 13.15 & 0.05 & 0.13 & 77.0     & 0.0073 &
& 4 \\
       & 4.89 & 33.7  & ---   & 13.15 & 0.05 & 0.13 & $\infty$ & 0.0071 &
& 4 \\
N5796  & 4.79 & 26.4  & 21.09 & ---   & ---  & ---  & ---      & 0.0189 &
$\setminus$ & \\
       & 4.70 & 25.6  & ---   & 13.70 & 0.04 & 0.70 & 92.6     & 0.0195 &
& 4 \\
       & 5.25 & 29.5  & ---   & 14.71 & 0.24 & 0.51 & $\infty$ & 0.0160 &
& \\
N5831  & 4.72 & 25.5  & 21.08 & ---   & ---  & ---  & ---      & 0.0038 &
$\setminus$ & \\
       & 4.77 & 25.9  & ---   & 13.66 & (0.04) & 0.30 & 77.8     & 0.0038 &
& \\
       & 4.72 & 25.5  & ---   & 13.01 & (0.01) & 0.00 & $\infty$ & 0.0039 &
& \\
N5845* & 2.74 & 4.57  & 18.65 & ---   & ---  & ---  & ---      & 0.0102 &
$\setminus$ & 3 \\
       & 2.88 & 4.36  & ---   & 15.79 & 0.68 & 0.58 & 6.74     & 0.0066 &
& \\
       & 2.82 & 4.44  & ---   & 15.64 & 0.59 & 0.57 & $\infty$ & 0.0066 &
& \\
\enddata

\tablecomments{Structural parameters for fits to the major-axis
profiles in our sample.  For each galaxy, we list in the first row the
best S\'ersic fit ($n$, $r_{e}$, $I_{e}$) and in the next two rows the
best \bomba{} fits ($n$, $r_{e}$, $r_{b}$, $I_{b}$, $\gamma$, and
$\alpha$; $\alpha = \infty$ is the sharp-transition version of the
\bomba{} model).  When $r_{b}$ is listed in parentheses, then its
value is $<$ the semi-major axis of the second innermost valid data
point; consequently, the corresponding power-law region is poorly
defined or meaningless.  The criteria for assigning galaxies to the
different categories (core, possible core, S\'ersic) are discussed in
the text.  Col.~(1): Galaxy name.  Cols.~(2)--(8): Best-fit parameters
of the S\'ersic and \bomba{} models (Eqs.~2, \ref{alphafree},
\ref{alpha-infty}).  The break radius $r_b$ and the effective radius
$r_e$ are in arcsec; $I_e$ and $I_b$ are in mag arcsec$^{-2}$
(observed values; no corrections for Galactic extinction or
cosmological effects have been made).  Col~(9): Reduced--$\chi^2$
values for the fits.  Col.~(10): Original \textit{HST} inner profile
classification from Nuker-law fits, from \citet{lauer95} or
\citet{rest01}; see Table~1.  Col.~(11): Notes --- 1 = inner
parameters ($r_{b}$, $\gamma$) dubious due to low value of $\alpha$; 2
= faint nuclear disk distorts profile; 3 = bright nuclear disk
distorts profile; 4 = $r_{b}$ of indicated fit is between second and
third data points of profile.}

\end{deluxetable}

\subsection{Core Galaxies} 

\begin{figure}
\begin{center}
\includegraphics[scale=0.9]{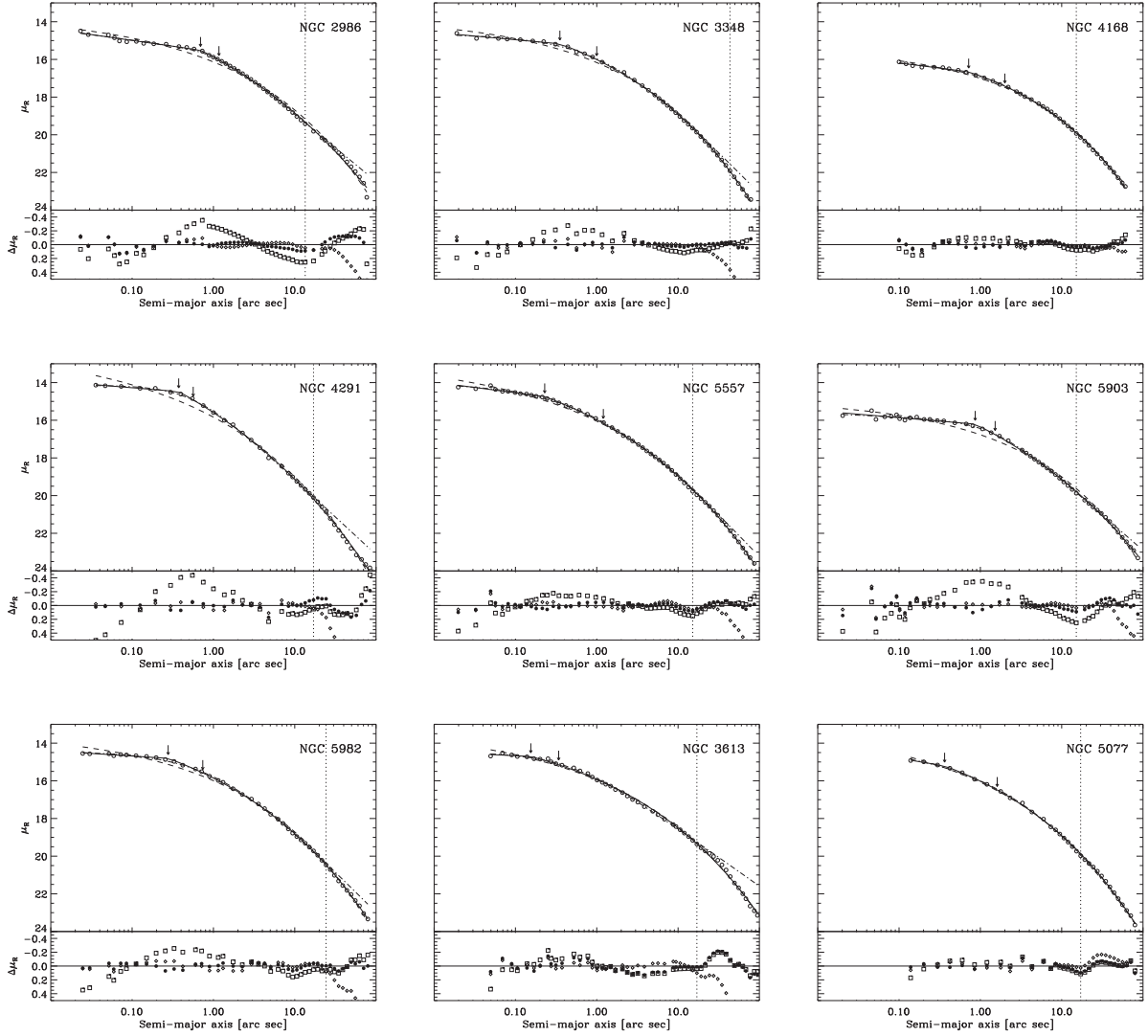}
\end{center}

\caption{Fits to the surface-brightness profiles (open circles) of
core galaxies.  For each galaxy, we show the best-fitting S\'ersic
(dashed line) and \bomba{} (solid line; $\alpha = \infty$ version)
models.  We also show the best-fitting Nuker-law profiles (dot-dashed
line), \textit{fit to the PC part of the profile only}; the outer
radius of the Nuker-law fits is marked by the vertical dotted line. 
Also shown are the residuals for each fit: S\'ersic (open squares),
\bomba{} (filled circles), and Nuker (small diamonds).  Finally, the
break radii of the \bomba{} (heavy arrow) and Nuker-law (light arrow)
fits are indicated.  In cases where the break radii of our Nuker-law
fits differ significantly from the published fits of \citet{rest01},
we indicate the \textit{published} value.}\label{fig:core-fits}

\end{figure}

Figure~\ref{fig:core-fits} shows the profiles and fits for the
galaxies we classify as ``core'' or ``possible core.''  Notice that
the pattern of the S\'ersic-fit residuals for these profiles match the
pattern in Figure~\ref{fig:core-residuals}: this is excellent
(qualitative) evidence for genuine cores in these galaxies.  As can be
seen, fitting the profiles with the \bomba{} model largely eliminates
these residuals.  Table~\ref{tab:fits} shows, in turn, that the
\bomba{} fits are significantly better, in a more quantitative,
statistical sense, than the S\'ersic fits for all but the two
``possible core'' galaxies: reduced chi-square values for S\'ersic
fits are larger by factors of $\sim 3$--15.

In general, we reproduce the core classifications of Rest et al.\
quite well, while finding that one of their ``intermediate'' galaxies
(NGC~5557) is actually a core galaxy.  We classify two galaxies,
NGC~3613 and NGC~5077, as ``possible core'' galaxies.  This is because
while the \bomba{} fits are better than the S\'ersic fits, they are
not significantly so: \chinu(CS) $< 2 \chinu($S\'ersic).  The patterns
of the S\'ersic-fit residuals for these galaxies in
Figure~\ref{fig:core-fits} do suggest possible core profiles, but
again this is not strong enough to be convincing.  In addition, the
break radii from the \bomba{} fits are near the inner limits of the
data; for NGC~3613, $r_{b} < 0\farcs16$, the nominal resolution
limit of the Nuker team's core definition.  For both galaxies, data at
smaller radii are needed to really confirm (or deny) the apparent
cores\footnote{\citet{rest01} noted an edge-on nuclear disk in the 
inner arc second of NGC~3613, which might explain some of the 
ambiguity if it is helping to mask a core, or producing a core-like 
break in the profile.}.

Table~\ref{tab:fits} includes the parameters and \chinu{} values for
fits using both variants of the \bomba{} model: free $\alpha$ and
$\alpha = \infty$ (sharp transition between power-law and S\'ersic
regimes).  By comparing the \chinu{} values for the core-galaxy and
possible-core fits, we can see that in most cases the $\alpha$ = free
fit is only marginally better than the $\alpha = \infty$ fit (see also
Column~4 of Table~\ref{tab:rms}).  As we suggested in
Section~\ref{sec:models}, the $\alpha = \infty$ model generally
provides just as good a fit as the free-$\alpha$ version, while having
one less free parameter \textit{and} having parameters values (e.g.,
$\gamma$) which better describe the modeled profile.

There is only one galaxy (NGC~4168) where the free-$\alpha$ fit is
significantly different, in terms of parameter values, from the the
$\alpha =\infty$ fit.  We suspect this difference is probably due to
the free-$\alpha$ model being better able to fit noise or extra
components in the profile, rather than being, e.g., an indication of a
core with a genuinely broad transition region.  First, there is
filamentary dust in the nuclear region \citep{rest01}, which produces
strong variations in the ellipse fits (Rest et al.\ and our
Figure~\ref{fig:efits}).  Second, the $\alpha = \infty$ break radius
($0\farcs72$, Table~\ref{tab:fits}) matches the apparent break in the
profile much better than the free-$\alpha$ value (3$\farcs15$), as can
be seen in Figure~\ref{fig:core-fits}.  Third, the S\'ersic index for
the $\alpha = \infty$ fit ($n = 3.1$) is more reasonable than the
free-$\alpha$ value ($n = 7.5$) for an intermediate-luminosity galaxy
\citep[see, e.g., Figure~10 of][]{graham-guzman03}.  Finally, the rms
residual values for both fits in the nuclear region
(Table~\ref{tab:rms}) are identical, which tells us that the
free-$\alpha$ fit does not provide a significantly better description
of the core.  For these reasons, we do not think the free-$\alpha$ fit
is genuinely better, and we prefer the $\alpha = \infty$ fit for
reasons of parsimony.

Finally, how do our \bomba{} fits compare with Nuker-law fits in terms
of reproducing the observed profiles?  Table~\ref{tab:rms} compares
rms residuals for the parts of the profile originally extracted from
the PC chip of WF/PC1 or WFPC2 and fit with the Nuker law by
\citet{byun96} and \citet{rest01}.  We remind the reader that the
\bomba{} fit is to the \textit{entire} profile, while the Nuker-law
fits are to the PC part of the profile only.  Thus, the Nuker-law fit
for NGC~3348, for example, is to semi-major axis $a =
0\farcs02$--14\farcs5, while the \bomba{} fit(s) are to $a =
0\farcs02$--78\farcs5; but the rms residuals are determined for the
same $a = 0\farcs02$-14\farcs5 region in both cases.

\begin{deluxetable}{cccccc}
\tabletypesize{\scriptsize}
\tablecolumns{6}
\tablewidth{0pc}
\tablecaption{Residuals of Fits in the Inner Region of
Galaxy Profiles\label{tab:rms}}
\tablehead{Galaxy & Profile ranges & S\'ersic rms & CS rms &
Nuker-law rms & Notes \\
(1)    &     (2)         &  (3)  &     (4)     & (5)   & (6)}
\startdata
\cutinhead{Core Galaxies} \\
N2986  & 0.02--14.4/76.9 & 0.19  & 0.054/0.054 & 0.053 &      \\
N3348  & 0.02--14.5/78.5 & 0.14  & 0.040/0.042 & 0.047 &      \\
N4168  & 0.10--14.7/60.7 & 0.073 & 0.037/0.037 & 0.037 &      \\
N4291  & 0.04--17.4/84.0 & 0.24  & 0.050/0.053 & 0.044 &      \\
N5557  & 0.02--14.6/86.5 & 0.12  & 0.041/0.050 & 0.044 &      \\
N5903  & 0.02--16.2/86.5 & 0.20  & 0.073/0.079 & 0.069 &      \\
N5982  & 0.03--17.0/79.2 & 0.15  & 0.030/0.037 & 0.044 &      \\
\cutinhead{Possible Core Galaxies} \\
N3613  & 0.05--18.4/94.5 & 0.11  & 0.068/0.070 & 0.049 &      \\
N5077  & 0.14--17.1/79.6 & 0.072 & 0.045/0.047 & 0.041 &      \\
\cutinhead{S\'ersic Galaxies} \\
N1426  & 0.35--10.2/81.6 & 0.041 & 0.030/0.031 & 0.015 &  1   \\
N1700  & 0.13--10.2/62.5 & 0.061 & 0.061/0.061 & 0.028 &  2   \\
N2634  & 0.10--13.7/55.5 & 0.066 & 0.046/0.046 & 0.027 &      \\
N2872  & 0.39--14.6/49.3 & 0.045 & 0.045/0.045 & 0.028 &  2   \\
N3078  & 0.63--16.7/79.2 & 0.025 & 0.025/0.025 & 0.015 &  2   \\
N4458  & 0.18--1.45/68.2 & 0.16  & 0.16/0.16   & 0.045 &  1,2 \\
N4478  & 0.02--14.9/70.3 & 0.23  & 0.23/0.15   & 0.11  &  1,3 \\
N5017  & 0.33--15.2/55.5 & 0.080 & 0.080/0.080 & 0.027 &  2   \\
N5576  & 0.02--16.0/77.5 & 0.073 & 0.063/0.064 & 0.046 &      \\
N5796  & 0.02--12.7/76.9 & 0.15  & 0.15/0.15   & 0.14  &  3   \\
N5831  & 0.02--14.9/68.3 & 0.061 & 0.057/0.061 & 0.053 &  3   \\
N5845  & 0.02--10.2/39.0 & 0.097 & 0.060/0.062 & 0.064 &  1   \\
\enddata

\tablecomments{Comparison of rms residuals for various fits in the
inner region (defined as that region fit with the Nuker-law for each
galaxy in Byun et al.~1996 or Rest et al.~2001).  The Nuker-law rms is
from our fit to the corresponding region; the S\'ersic and \bomba{}
(CS) rms are from our fits to the \textit{entire} profile, with the
residuals calculated in the inner region only.  Col.~(1): Galaxy name. 
Col.~(2): Fitted regions of profile (semi-major axis, in arc seconds). 
The first range is the ``inner region'' (fit with Nuker law), followed
by outer limit of the S\'ersic and \bomba{} fits.  Col.~(3): rms
residuals, in magnitudes, of S\'ersic fit, calculated in Nuker-law fit
region.  Col.~(4): same as (3), but for the \bomba{} fits --- first
number is for free-$\alpha$ version, second is for $\alpha = \infty$. 
Col.~(5): rms residuals of Nuker-law fit.  Col.~(6) Notes: 1 = nuclear
disk; 2 = both \bomba{} fits reproduce S\'ersic fit; 3 $\alpha =
\infty$ \bomba{} fit reproduces S\'ersic fit.}

\end{deluxetable}

For the core galaxies, the \bomba{} fit residuals in the PC region are
never more than 20\% larger than the Nuker-law residuals; the mean
excess is only 3\%, and for three of the seven galaxies, the \bomba{}
residuals are equal to or \textit{less than} the Nuker-law residuals. 
This is rather astonishing, given that the \bomba{} fit is constrained
to fit the profiles out to $\sim 5$ times further in radius while
still having approximately the same number of parameters (exactly the
same, in the case of the $\alpha = \infty$ \bomba{} model).  Casual
inspection of Figure~\ref{fig:core-fits} shows that the Nuker-law fits
become much worse than the \bomba{} fits outside the PC part of the
profile, as might be expected.  We also note that the
\textit{parameter} $\gamma$ from our $\alpha = \infty$ \bomba{} fits
is usually a closer match to the \textit{observed} slope ($\gammap$,
evaluated at $r = 0\farcs1$, from \nocite{rest01}Rest et al.\ 2001)
than is the Nuker-law parameter $\gamma$; see
Table~\ref{tab:core-params}.

\begin{deluxetable}{cccccccc}
\tabletypesize{\scriptsize}
\tablecolumns{8}
\tablewidth{0pc}
\tablecaption{Comparison of \Bomba{} and Nuker Parameters for 
Cores\label{tab:core-params}}
\tablehead{ Galaxy  & $I_b$(CS) & $r_b$(CS) & $\gamma$(CS) &
$I_b$(Nuk) & $r_b$(Nuk) & $\gamma$(Nuk) & $\gammap$ \\
              (1)    & (2)  &  (3) & (4)  & (5)& (6) & (7) & (8)}
\startdata
N2986  & 15.5 &  0.69/97  & 0.25   & 16.1 &  1.24/174 & 0.18  & 0.20 \\
N3348  & 15.2 &  0.35/70  & 0.16   & 16.0 &  0.99/198 & 0.09  & 0.18 \\
N3613  & 14.7 &  0.15/21  & 0.09   & 15.1 &  0.34/48  & 0.04  & 0.17 \\
N4168  & 16.7 &  0.72/108 & 0.22   & 17.5 &  2.02/303 & 0.17  & 0.19 \\
N4291  & 14.5 &  0.37/47  & 0.14   & 15.1 &  0.60/76  & 0.00  & 0.13 \\
N5077  & 15.2 &  0.36/62  & 0.29   & 16.5 &  1.61/279 & 0.23  & 0.30 \\
N5557  & 14.7 &  0.23/51  & 0.23   & 16.2 &  1.21/269 & 0.14  & 0.33 \\
N5903  & 16.2 &  0.86/141 & 0.15   & 16.8 &  1.59/262 & 0.13  & 0.14 \\
N5982  & 14.8 &  0.28/57  & 0.11   & 15.6 &  0.74/151 & 0.00  & 0.18 \\
\enddata

\tablecomments{Comparison of core parameters obtained from \bomba{}
(CS) and Nuker-law (Nuk) fits to the core galaxies.  The break radii
$r_{b}$ are in arc seconds/parsecs; $R$-band surface brightness at the
break radius is in mag arcsec$^{-2}$.  We use the $\alpha = \infty$
(sharp-transition) version of the \bomba{} model for the CS values;
the Nuker-law values and the slope at $r = 0.1\arcsec$ ($\gammap$) are
taken from the original fits in \citet{rest01}.}

\end{deluxetable}

\subsection{S\'ersic Galaxies} 

\begin{figure}
\begin{center}
\includegraphics[scale=0.85]{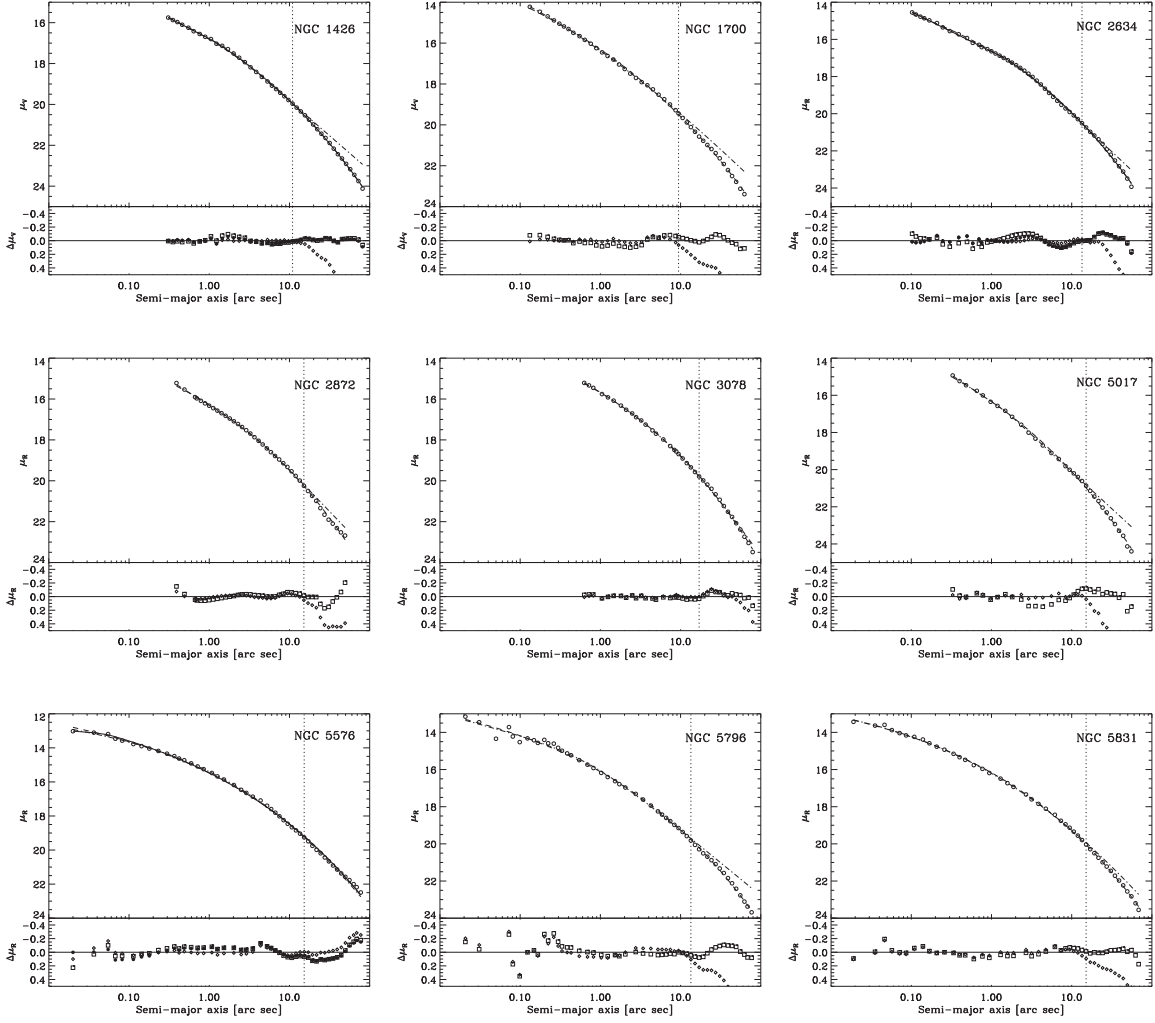}
\end{center}

\caption{As for Figure~\ref{fig:core-fits}, but showing fits to the
surface-brightness profiles of S\'ersic (i.e., non-core) galaxies.  In
several cases, the best \bomba{} fit is \textit{identical} to the best
S\'ersic fit, so just the S\'ersic and Nuker-law fits are
show.}\label{fig:sersic-fits}

\end{figure}

\begin{figure}
\begin{center}
\includegraphics[scale=0.85]{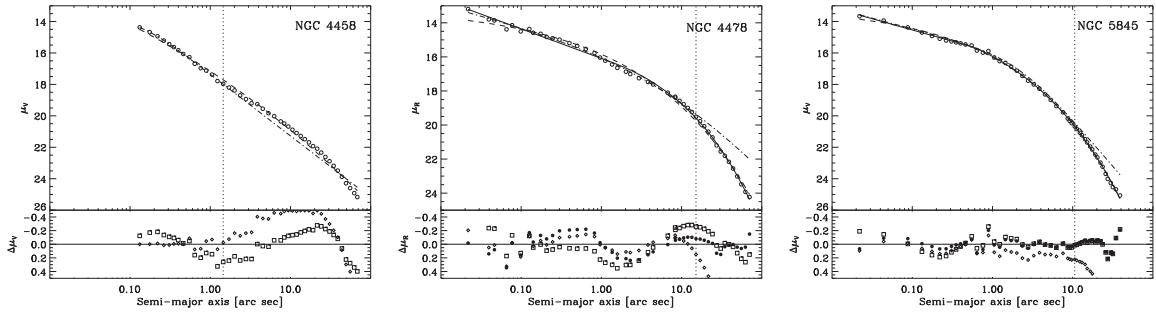}
\caption{As for Figure~\ref{fig:sersic-fits}, but showing fits for 
galaxies with prominent nuclear disks.}\label{fig:sersic-fits-bad}
\end{center}

\end{figure}

The remaining twelve galaxies (Figures~\ref{fig:sersic-fits} and
\ref{fig:sersic-fits-bad}) are those for which there is no clear
evidence for a core: the residuals of the S\'ersic fits do not display
the characteristic ``core pattern'' (Figure~\ref{fig:core-residuals}),
and the \bomba{} fits are not significantly better in terms of
\chinu{}.  In fact, for seven of these twelve galaxies one or both of
the best \bomba{} ($\alpha$ free or $\alpha = \infty$) fits
\textit{reproduces} the best S\'ersic fit: the $n$ and $r_{e}$
parameters are identical, and the \bomba{} break radius $r_{b} <$ the
innermost data point.  \Bomba{} fits of this nature are clear evidence
that these galaxies' profiles are well described by pure S\'ersic
profiles.  For another four of the galaxies, the $n$ and $r_{e}$
parameters differ by less than 5\% between the \bomba{} and S\'ersic
fits, and so the pure S\'ersic profile is also preferred for reasons
of simplicity.

All twelve of these galaxies were previously classified as power-law
galaxies by \citet{byun96} or \citet{rest01}, based on their Nuker-law
fits.  A comparison of the residuals (Table~\ref{tab:rms}) shows that
the Nuker law does fit the inner (PC) profiles slightly better,
though, as Figures~\ref{fig:sersic-fits} and \ref{fig:sersic-fits-bad}
show, the Nuker-law residuals are always worse --- usually much worse
--- at larger radii.  It is not too surprising that a fit using five
parameters (the Nuker law), restricted to the inner 10--17\arcsec,
does better in that region than a fit using only three parameters
which also fits the profile out to 3--8 times further in radius. 
Nonetheless, for six of these galaxies, the (inner) S\'ersic-fit rms
residuals are $<$ 2 times the Nuker-law residuals, and for only one
galaxy are the S\'ersic residuals $>$ 3 times the Nuker-law residuals. 
As we discuss below, the strongest discrepancies are probably due to
extra components such as nuclear disks.

There are five power-law galaxies where the Nuker fit is clearly
better (in the inner region) --- NGC~1426, 2634, 4458, 4478, and 5017. 
In four of these galaxies (NGC~1426, 4458, 4478, and 5845), there is
clear evidence for a luminous nuclear disk (see
Figure~\ref{fig:ndisks} and the ellipse fits in
Appendix~\ref{app:figs}), with the break radius in the Nuker-law fits
(and some of the \bomba{} fits) occurring close to the point of maximum
ellipticity associated with the nuclear disks.  The distortions
created by the nuclear disks in NGC~4458 and NGC~4478 are so strong
--- producing the largest residuals of any of the galaxies --- that we
do not consider the S\'ersic fits to be reliable.  A similarly strong
nuclear disk (combined with a dust disk) is found in NGC~5845
\citep[e.g.,][]{quillen00}, so the S\'ersic fit there may not be
reliable either, although the Nuker-law fit is not dramatically
better.  There is evidence for a slight break in NGC~2634's
surface-brightness profile at $a \sim 2\arcsec$, though there is no
accompanying signature in the ellipse fits --- perhaps a face-on
nuclear disk?  NGC~5017 is also somewhat mysterious, but the fact that
the \bomba{} fits reproduce the S\'ersic fit (Table~\ref{tab:fits})
shows that this is not a core galaxy, and we tentatively include it
with the S\'ersic galaxies.

We note that the residuals for \textit{all} of the fits to NGC~5796 
are large, but this is clearly attributable to the noise in the 
profile at $a < 0\farcs2$.

\section{Discussion}
\label{sec:discuss}

We conclude that most, if not all, ``power-law'' ellipticals are
probably best understood as having S\'ersic profiles --- modulo extra
components such as nuclear star clusters, nuclear disks, etc.\ --- into
the limits of resolution (or limits imposed by dust).  As discussed in
Section~2, this is consistent with an overall trend for elliptical
galaxies: low- and intermediate-luminosity ellipticals have pure
S\'ersic profiles (plus optional nuclear disks, clusters, and point
sources), and distinct cores appear in high-luminosity systems as
deviations from the outer S\'ersic profile.
\nocite{graham-guzman03}(Graham \& Guzm\'an 2003 combine measurements
for a large set of elliptical galaxies, including dwarf ellipticals, to
make this argument in more detail.)  Moreover, for power-law galaxies,
we get excellent fits using a model with fewer parameters, all of which
are physically meaningful (i.e., correlate with other galaxy
parameters).  These fits work for the \textit{entire} profile,
\textit{unlike} the Nuker law, yet are as good a fit in the region
where the Nuker law is usually used.

The term ``power-law galaxy'' is thus somewhat misleading, since it
suggests that the nuclear profile is adequately described by a single
power-law, which is probably different from the outer profile.  While
this is an appealingly simple description for modeling purposes, our
results strongly suggest that this is not accurate.  Instead,
elliptical galaxy profiles have logarithmic slopes which continuously
decrease as $r \rightarrow 0$.  Figure~11 of \citet{lauer95}, which
presents representative examples of ``power-law'' profiles, supports
this argument: even the galaxy which is closest to a perfect power-law,
NGC~1700, shows a systematic deviation from a power-law --- steeper at
larger radii, shallower at smaller radii --- as expected for a S\'ersic
profile; see Figure~\ref{fig:sersic-fits}.  (This is not the case for
the central cusps of \textit{core} galaxies; their Figure~7.)

The ``intermediate'' galaxies reported by \citet{rest01} and
\citet{ravindranath01} are probably a consequence of Nuker-law fits
applied to this overall elliptical-galaxy trend.  Lower-luminosity
``intermediate'' galaxies are most likely S\'ersic galaxies with low
values of $n$ (and hence $<\gamma>$ in the range 0.3--0.5; see
Fig.~\ref{fig:de-gamma}).  At higher luminosities, core galaxies can
appear to have $\gamma > 0.3$ if the core is not adequately resolved
(either due to distance or to inner truncation of the profile by,
e.g., dust).  (We \textit{do} classify two galaxies in our sample as
``possible core'' galaxies, but these are clearly cases of inadequate
resolution.)

Although we have not yet attempted to model the complete profiles of
\textit{bulges}, it is reasonable to extend our results to them. 
\citet{balcells03} have already done this for a sample of early-type
bulges in the near-IR, using NICMOS data in conjunction with
ground-based imaging.  They find that the complete bulge profiles,
after accounting for the presence of the outer disk, can be well
modeled by S\'ersic profiles, plus optional nuclear components
(corresponding to, e.g., nuclear star clusters or point sources). 
This is in excellent agreement with our hypothesis that the profiles
of lower-luminosity ellipticals and bulges are fundamentally S\'ersic
profiles, and promises to resolve a number of ambiguities and
``dichotomies'' reported in the literature.  For example,
\citet{carollo97} and \citet{seigar02} argue for a dichotomy between
$R^{1/4}$ and exponential bulges, with the latter having low $\gamma$
in contrast to the high $\gamma$ of $R^{1/4}$ bulges and
moderate-luminosity ellipticals.  This is naturally explained if most
bulges actually have S\'ersic profiles (as is well supported by a
number of studies) \textit{and} if these S\'ersic profiles extend into
the nuclear region.  The division between $R^{1/4}$ (S\'ersic index $n
= 4$) and exponential ($n = 1$) bulges is probably an artificial one,
given that bulges in reality show a range of values of $n$.  But as
\nocite{paper1}Paper~I shows, bulges with larger $n$ will have higher
values of $\gamma$ than bulges with low $n$.  Thus, ``$R^{1/4}$''
bulges (higher $n$) will exhibit larger values of $\gamma$ than
``exponential'' (lower $n$) bulges.  Since bulge $n$ decreases along
the Hubble sequence, the trend of decreasing $\gamma$ with Hubble type
noted by \citet[][ their Fig.~3]{seigar02} follows as well.

In retrospect, we can see that most of the early \textit{HST} studies
of galaxy centers, and some of the more recent ones
\citep[e.g.,][]{rest01,ravindranath01}, have focused on relatively
high-luminosity systems.  These samples thus included a mix of
S\'ersic galaxies with high $n$ values and genuine core galaxies,
making a distinction between core and ``power-law'' galaxies based
purely on $\gamma$ feasible.  More recent studies aimed at
low-luminosity systems \citep[e.g.,][]{carollo98,stiavelli01,seigar02}
have since uncovered evidence for the low-$n$--low-$\gamma$,
high-$n$--high-$\gamma$ trend that pure S\'ersic profiles generate,
and thus show that discriminating core galaxies purely by $\gamma$ is
problematic at best.

\subsection{Core Identifications and Core Parameters} 
\label{sec:core-params}

We find that most of the previously identified ``core'' galaxies in
our sample \textit{do} have distinct cores with shallow, power-law
cusps.  These cores stand out as downward deviations from the outer
S\'ersic profiles.  Fitting with the \bomba{} model provides a more
natural, less ambiguous definition for ``true'' cores, without the
possibility of misclassifying low-$n$ S\'ersic profiles as cores.  We
are also able to re-classify one of the ``intermediate'' galaxies
(NGC~5557) of \citet{rest01} as a core galaxy.  The two ambiguous
galaxies --- NGC~3613 and NGC~5077 --- are simply cases where the
apparent break radius is very close to the inner limits of the data. 
For NGC~3613, this is because the apparent core is close to the
resolution limit (in fact, $r_{b}$ from the \bomba{} fits is $<
0\farcs16$ and thus smaller than the suggested resolution-based
limit of \nocite{faber97}Faber et al.\ 1997).  For NGC~5077, on
the other hand, \citet{rest01} clipped their data at $r = 0\farcs1$
because of an apparent nuclear excess at smaller radii.  A future fit
including data at smaller radii and using an extra nuclear component
to account for this excess, may help determine if NGC~5077 truly
possesses a core.

While our overall agreement with the core/non-core classifications of
\citet{lauer95} and \citet{rest01} is quite good for the galaxies we
analyze, \textit{we find that Nuker-law fits systematically
overestimate the size of the cores}: our break radii are $\sim
1.5$--4.5 times smaller in size than the break radii from the
published Nuker-law fits.  Consequently, $\mu_{b}$ values are brighter
as well.  We also find consistently higher values of $\gamma$, though
the difference is not as dramatic (see Table~\ref{tab:core-params} and
Figure~\ref{fig:core-core}).  This is in excellent agreement with the
arguments of \nocite{paper1,paper3}Papers~I and III: \textit{all}
Nuker-law parameters are sensitive to the radial size of the region
where the fit is made.  All parameters of the Nuker model, including
$\gamma$ and $r_{b}$, must be adjusted in order to fit both the core
\textit{and} the (non--power-law) part of the profile outside, with
its intrinsic (S\'ersic) curvature.  Table~\ref{tab:core-params} shows
that, on average, the \bomba{} values of $\gamma$ match the
\textit{observed} core slope $\gammap$ (as determined by
\citet{rest01}) better than the Nuker-law values do.

The currently favored theory for core formation is the ejection of
core stars by 3-body encounters with a decaying black hole binary
formed following a merger of two galaxies with central supermassive
black holes.  Various calculations
\citep{ebisuzaki91,quinlan97,milosavljevic01} have estimated the
stellar mass ejected during this process (\mej), and generally find it
to be $\sim \mbh$, where \mbh{} is the mass of the resulting central
black hole formed by the (assumed) coalescence of the binary. 
However, attempts to test these predictions by estimating \mej{} from
observed cores and comparing it with various estimates of \mbh{}
consistently produce values of $\mej > \mbh$.  \citet{faber97} found
$\mej = 3.5$--6.4 \mbh; using more accurate estimates of \mbh,
\citet{milosavljevic01} found $\mej \approx 1$--20 \mbh. 
\citet{ravindranath02} used the prescription for \mej{} of
Milosavljevi\'{c} \& Merritt and a much larger data set; they found
$\mej \approx 2$--20 \mbh{} at the low-mass end ($\mbh \sim 10^{8}
M_{\sun}$), while at the high-mass end ($\mbh \sim 10^{9} M_{\sun}$)
$\mej \approx 6$--25 \mbh.  Even considering only the galaxies with
\textit{measured} \mbh, $\mej/\mbh \approx$ 4--13.  Milosavljevi\'{c}
\& Merritt pointed out that the \textit{total} ejected mass should
increase with the number of mergers, but the observed ratios still
seem high, particularly at the low-mass end, where there have
presumably been fewer mergers.

All of the studies cited above used parameters from Nuker-law fits to
estimate \mej.  Since the estimated \mej{} scales with $r_{b}$ --- in
the parameterization introduced by \citet{milosavljevic01} and used by
Ravindranath et al.\ (2002), $\mej \propto r_{b}$ --- overestimating
$r_{b}$ will naturally overestimate \mej.  Thus at least some of the
discrepancy between observed and predicted $\mej/\mbh$ is probably due
to the tendency of Nuker-law fits to overestimate $r_{b}$, as we have
found.  Assuming that the core radii from \bomba{} fits are typically
$\sim 2$--4 times smaller than the Nuker-law values, as is the case
for our sample, $\mej/\mbh$ values should go down by comparable
factors, which would put them in better agreement with the theoretical
predictions.

One of our core galaxies (NGC~4291) was noted by
\citet{ravindranath01} for possibly having an isothermal core (with
$\gamma = 0$), on the basis of their Nuker-law fits to a NICMOS image. 
The Nuker-law fit in \citet{rest01} to the WFPC2 profile also has
$\gamma = 0.0$, which might seem to strengthen the case for an
isothermal core.  However, we find $\gamma = 0.14$ from our \bomba{}
fit, which agrees very well with $\gammap = 0.13$ determined by Rest
et al.  So the core of NGC~4291 is probably \textit{not} isothermal.

In Figure~\ref{fig:core-global} we show the relation between the core
and the global properties of the galaxies in our sample.  We also
indicate the upper limits on possible core radii for the S\'ersic
galaxies, based on the radii of the innermost valid data.  For those
galaxies where a clear core has been measured, we find that the
relation between the break radius and the effective radius is
approximately given by $r_b = 0.014 r_e$.  This is a factor of two
smaller than the relation found by \citet{faber97}, consistent with
our finding that fitting with the Nuker law tends to overestimate core
sizes.

There is a suggestion of a weak trend of $r_{b}$ increasing with galaxy
luminosity, which would be in agreement with what Faber et al.\ found
\citep[see also][]{laine03}, but for our sample this ``trend'' is
anchored by only two points, so it is dubious.  Unfortunately, the
narrow magnitude range spanned by the core galaxies in our sample
($\lesssim 1.5$ mag) precludes a proper test of the magnitude-$r_{b}$
relation reported Faber et al., which is based on galaxies spanning
$\gtrsim 3$ mag (and the composite trend in Fig.~9 of Laine et al.\
spans almost 5 magnitudes).  There is no clear magnitude-related trend
in the \textit{ratio} of our $r_{b}$ measurements to the Nuker-law
measurements, which suggests that the magnitude-$r_{b}$ trend may be
unaffected by changes in $r_{b}$, except possibly in the scatter.
However, a proper evaluation of how the magnitude-$r_{b}$ relation is
affected by better measurements of $r_{b}$ must await \bomba{} fits to
a larger sample of core galaxies.  There is \textit{no} evidence for a
relationship between $n$ and $r_{b}$; this may be partly due to large
uncertainties in $n$ \nocite{caon93}(Caon et al.\ 1993 found typical
errors of $\sim 25$\% when fitting S\'ersic profiles).  Finally, we
find \textit{no} clear correlation between $\gamma$ and the global
properties of the core galaxies analyzed.  This is agreement with what
previous studies have found for core galaxies (e.g.,
\nocite{rest01}Rest et al.\ 2001, Figure~7;
\nocite{ravindranath01}Ravindranath et al.\ 2001, Figure~3;
\nocite{laine03}Laine et al.\ 2003, Figure~6; and the core galaxes in
Figure~\ref{fig:gamma-core} of this paper).

\subsection{Hidden Cores and the Core-Galaxy Fraction} 

An interesting point is to consider how well-resolved the underlying
profiles of the various galaxies actually are.  In several cases,
\citet{byun96} and \citet{rest01} excluded points at small radii from
their fits, usually due to the presence of significant nuclear dust or
a distinct nuclear component (e.g., a nuclear point source).  Thus,
not all of the profiles take full advantage of \textit{HST}
resolution.  While the nuclear components may include cases of nuclear
star clusters, which make discussions of the underlying stellar
profile ambiguous, the presence of dust means that some ``power-law''
(i.e., S\'ersic-profile) galaxies could have hidden cores.

If we divide the sample into two groups --- galaxies where the
innermost valid data point is at $r < 15$ pc (spatially well resolved
centers); and galaxies where the innermost valid point is at $r > 15$
pc (less well-resolved centers) --- we find that the \textit{less}
resolved galaxies are almost all\footnote{The exceptions are NGC~4168
(core) and NGC~5077 (possible core).} well fit using just the S\'ersic
model.  This suggests that at least some of the S\'ersic galaxies
could have ``hidden'' cores.  This is not a new argument, obviously,
as many authors have pointed out that ``power-law'' galaxies could
include unresolved cores --- but it is interesting to consider how
\textit{few} of the S\'ersic galaxies in our sample can really be
declared free of \textit{HST}-resolvable cores.  Of the 21 galaxies,
seven clearly have cores, two have possible cores (NGC~3613 and
NGC~5077, see Section~\ref{sec:core-params}), and only five (NGC~4478,
NGC~5576, NGC~5796, NGC~5831, and NGC~5845) are clearly free of
significant ($r_{b} > 5$ pc) cores.

So in the limited range of absolute magnitude spanned by our full
sample ($-18.3 \gtrsim M_{B} \gtrsim -21.4$), 33\% of the galaxies have
unambiguous, \textit{HST}-resolved cores; but this is clearly a lower
limit.  The core fraction rises to 43\% if we include the two possible
cases, and in principle could be as high as 76\%.  It is also
interesting to note that we can see in the absolute magnitudes a hint
of the well-known dichotomy between core and non-core galaxies
\citep[see, e.g., the discussion in ][]{rest01}, even in our limited
sample.  This can be seen in Figure~\ref{fig:core-global}, where the
five \textit{fully resolved} S\'ersic galaxies tend to be fainter than
the core galaxies; a Kolmogorov-Smirnov test gives a 95\% probability
that the two groups of galaxies come from different parent luminosity
distributions.

\section{Summary}
\label{sec:summary}

We have successfully fit the complete surface-brightness profiles of
19 out of 21 elliptical galaxies, from the \textit{HST}-resolved
central regions ($r \sim 0\farcs02$) out to $\sim$ twice the
half--light radius, using either: a) a pure S\'ersic profile; or b) a
``\bomba{}'' model consisting of an outer S\'ersic profile joined to
an inner power-law core.  The former fits correspond to so-called
``power-law'' galaxies, which are perhaps better described as
``S\'ersic galaxies,'' and the latter correspond to core galaxies.

The combined use of these two models lets us address the following 
questions:
\begin{enumerate}

\item \textit{How can we relate the central, HST-resolved part of the
galaxies' surface-brightness profiles to the outer regions?} We show
that most power-law ellipticals are well described at all radii by the
simple S\'ersic law (modulo any nuclear disks, etc.).  On the other
hand, core galaxies are extremely well fit with the \bomba{} model.  We
find little need for a significant transition region between the outer
(S\'ersic) part of the \bomba{} profile and the (power-law) core; any
such transition region is small compared to the size of the core.

\item \textit{Is there a dichotomy in nuclear profiles between low-
and high-luminosity bulges and ellipticals?} Some recent \textit{HST}
studies have suggested that the apparent trend seen in intermediate-
and high-luminosity bulges and ellipticals --- cores with shallow
logarithmic slopes in high-luminosity systems, steeper nuclear slopes
in lower-luminosity (``power-law'') systems --- breaks down at lower
luminosities, because fainter bulges and dwarf ellipticals have
shallow nuclear slopes.  We show that the power-law galaxies in our
sample have S\'ersic profiles that extend into the limits of
\textit{HST} resolution, with $n \sim 4$--6; this naturally explains
the steep nuclear slopes previously reported.  When combined with the
well-known correlation between $n$ and luminosity, we can see that
\citep[as argued by][]{graham-guzman03} the general trend is most
likely one of pure S\'ersic profiles (plus possible extra components
such as nuclear star clusters and disks), extending from
low-luminosity systems with low-$n$ S\'ersic profiles --- and thus
shallow nuclear slopes --- to high-luminosity systems with high-$n$
profiles and steeper nuclear slopes.  Only the high-luminosity
\textit{core} galaxies break the trend, due to the existence of the
cores themselves.

\item \textit{How can we unambiguously identify cores in galaxy
profiles?} As we demonstrate, the traditional definition of cores
using parameters from Nuker-law fits to galaxy profiles ($r_b \geq
0\farcs16$ and $\gamma < 0.3$) leads to the real possibility of
misclassifying galaxies with sufficiently shallow slopes (for example,
exponential profiles) as core galaxies.  \textit{We define core
galaxies as those possessing a well-resolved downward deviation from
the inward extrapolation of the outer (S\'ersic) profile}.  This
definition recovers previous core definitions for the high-luminosity
ellipticals in our sample, but is immune to the danger of identifying
exponential-like profiles as having cores.

\item \textit{How can we more accurately determine the structural
properties of cores?} As demonstrated in Paper~I, the Nuker law
requires a broad, smooth transition (low values of $\alpha$) between
its two power-law regimes in order to fit the inner profiles of core
and power-law galaxies, because this is the only way to reproduce the
observed curvature of actual galaxy profiles.  We find that this
causes the core-size measurements (i.e., the break radius) to be
overestimated by factors of 1.5--4.5 in comparison to the values
derived by using the \bomba{} model, which directly accounts for the
intrinsic curvature of galaxy profiles.  We also find that the
logarithmic slope $\gamma$ of the observed core is more accurately
recovered with the \bomba{} model.  Using the smaller values we find,
especially for $r_{b}$, should bring estimates of the ejected stellar
mass due to core formation more in line with theoretical predictions.

\end{enumerate}

\acknowledgements

We thank Linda S. Sparke for useful discussions.  This research is
based on observations made with the NASA/ESA Hubble Space Telescope,
obtained from the data archive at the Space Telescope Institute. 
STScI is operated by the association of Universities for Research in
Astronomy, Inc.\ under the NASA contract NAS 5-26555.

This research also made use of the Lyon-Meudon Extragalactic Database
(LEDA; http://leda.univ-lyon1.fr), and the NASA/IPAC Extragalactic
Database (NED), which is operated by the Jet Propulsion Laboratory,
California Institute of Technology, under contract with the National
Aeronautics and Space Administration.



\onecolumn

\begin{figure}
\begin{center}
\includegraphics[scale=0.9]{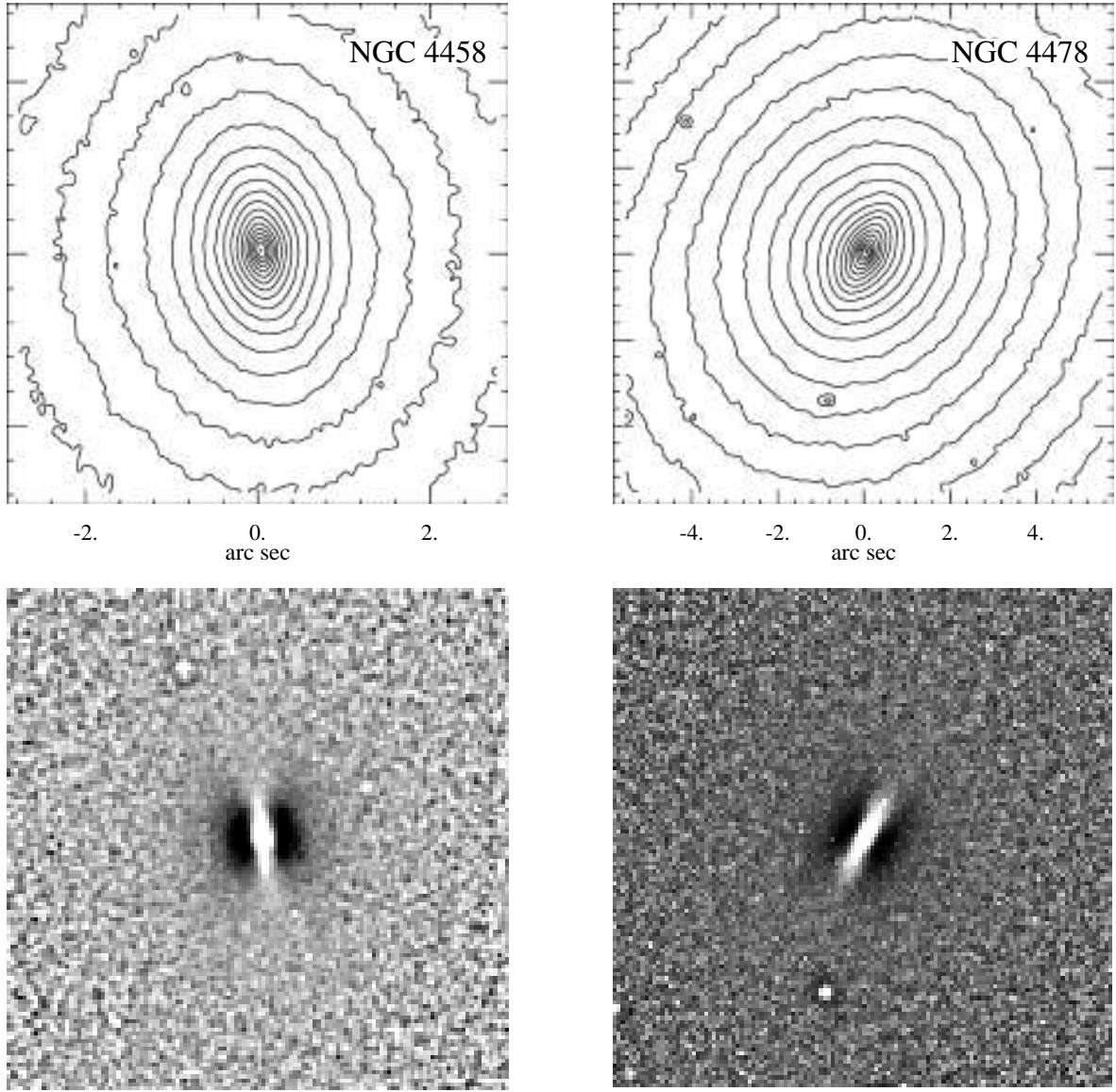}
\end{center}
\caption{Isophote contours (top) and unsharp masks (bottom) of PC
images of NGC~4458 and NGC~4478, showing the prominent nuclear disks
in each (see also the ellipse fits in Figure~\ref{fig:efits}).  These
nuclear disks introduce strong deviations from a pure S\'ersic models
in the surface brightness profiles.  \citep[Similar effects are
produced by the nuclear disk in NGC~5845;
see][]{quillen00}.}\label{fig:ndisks}

\end{figure}

\begin{figure}
\begin{center}
\includegraphics{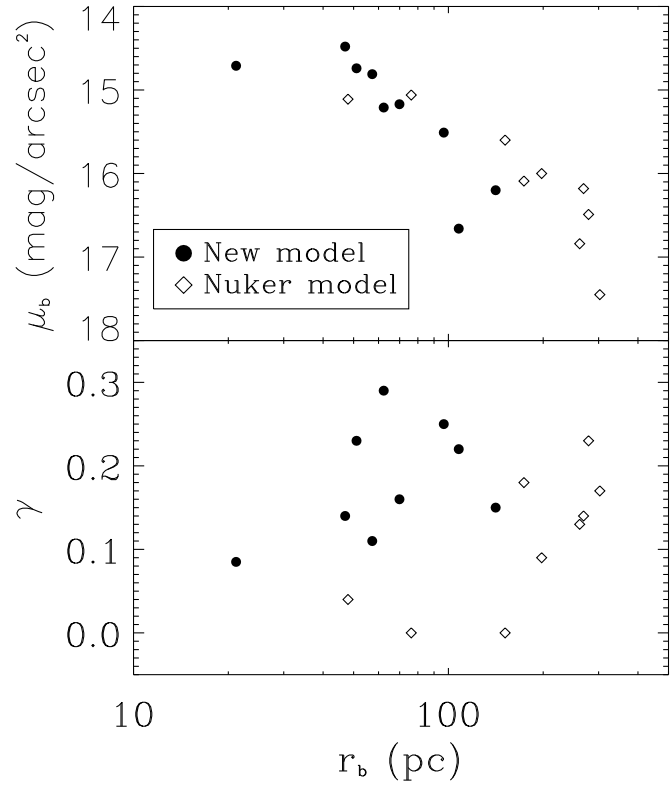}
\end{center}

\caption{Core properties (break radius $r_{b}$, inner logarithmic
slope $\gamma$, and surface brightness at the break radius $\mu_{b}$)
for the core galaxies in our sample.  Filled circles are our
measurements, using the \bomba{} fits; open circles are published
values from Nuker-law fits \citet{rest01}.}\label{fig:core-core}
\end{figure}

\begin{figure}
\begin{center}
\includegraphics{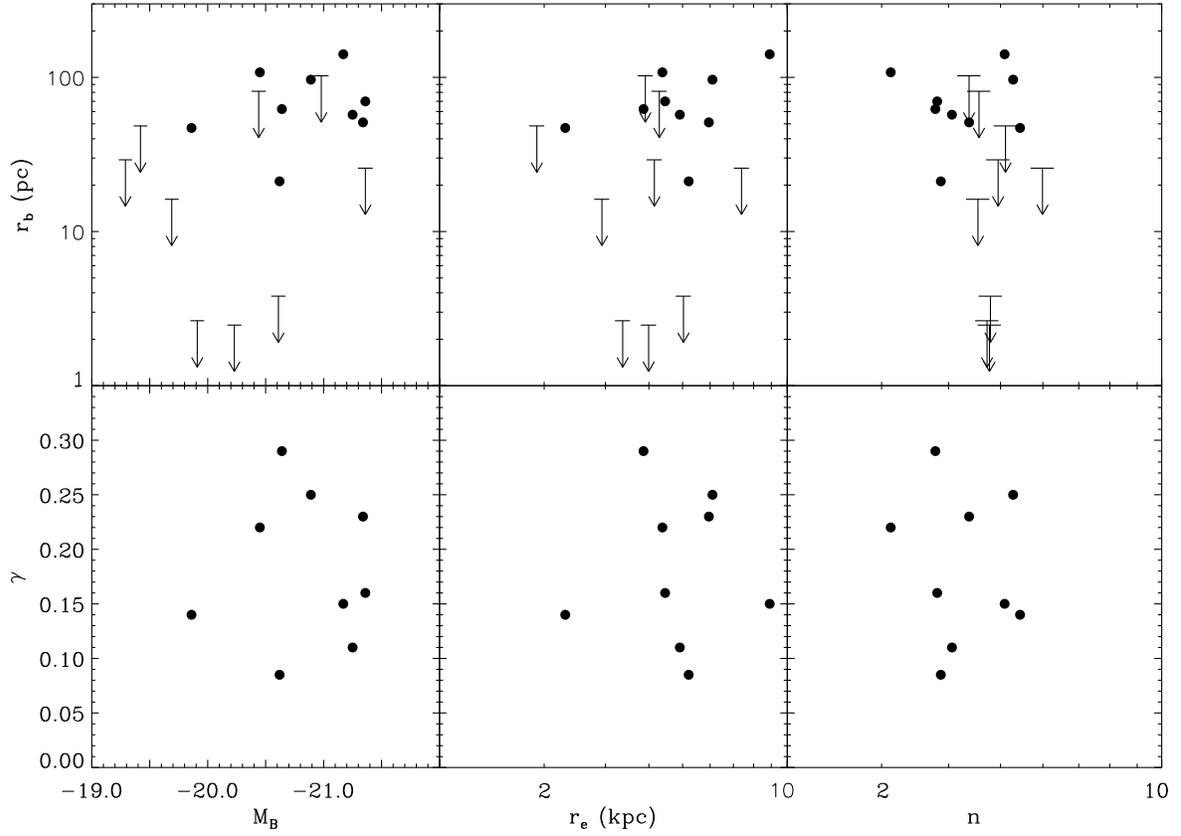}
\end{center}

\caption{Comparison of core properties (break radius $r_{b}$ and inner
logarithmic slope $\gamma$) and global properties for the core
galaxies in our sample.  The upper limits on possible break radii for
the S\'ersic galaxies (based on the innermost fitted data point; see
Column~8 of Table~\ref{tab:sample}) are indicated by the arrows. 
Three of the latter (NGC~4458, 4478, and 5845) have S\'ersic fits that
are distorted by bright nuclear disks --- see
Figures~\ref{fig:sersic-fits-bad} and \ref{fig:ndisks} --- so we do
not plot their $r_{e}$ and $n$ values.}\label{fig:core-global}
\end{figure}

\begin{figure}
\begin{center}
\includegraphics{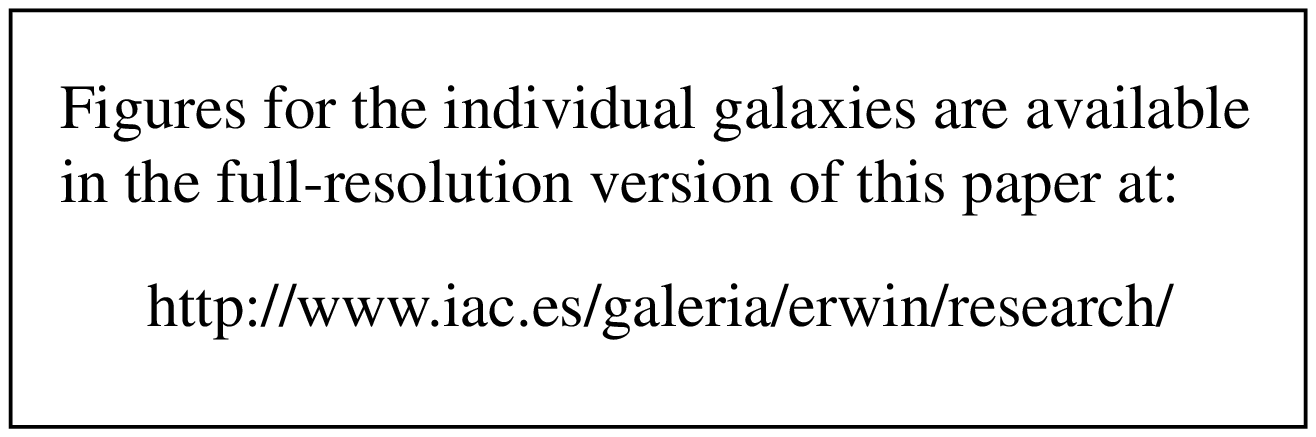}
\end{center}

\caption{Isophotes and ellipse fits for WFPC2 mosaic images of the 21
elliptical galaxies in our sample.  Contour plots of the isophotes
show the entire WFPC2 array; the coordinate axes are centered on the
galaxy nucleus.  Isophotes are logarithmically scaled and have been
smoothed with a 5-pixel-wide median filter prior to contouring. (Due 
to size constraints, these figures are only available in the 
full-resolution version of this paper, at:
\texttt{http://www.iac.es/galeria/erwin/research/}). }\label{fig:efits}
\end{figure}

\appendix

\section{Some Useful Mathematical Expressions Related to the \Bomba{} Model} 
\label{app:math}

\subsection{The Relation Between \Bomba{} and S\'ersic Effective
Radii}

In this section we want to prove the following identity:
\begin{equation}
b_n\bigg(\frac{1}{r_{es}}\bigg)^{1/n} = \;
b\bigg(\frac{1}{r_{e}}\bigg)^{1/n},
\end{equation}
where $r_{es}$ is the effective radius of the S\'ersic part of the
\bomba{} model, $r_{e}$ is the effective radius of the global \bomba{}
model, and $b_n$ and $b$ are the quantities introduced in order to give
to $r_e$ in the S\'ersic and \bomba{} model, respectively, the meaning
of effective radius.

Demonstration: Although the above relation can be proved for smooth
transitions between the S\'ersic regime and the power-law regime (i.e.,
$\alpha$ small), we will only show the demonstration for the sharpest
transition case ($\alpha \rightarrow \infty$).  The S\'ersic part of
the \bomba{} model is described using the following law:
\begin{equation}
I(r) \,=\, I(0) \, \exp[-b(r/r_{e})^{1/n}],
\label{sersextr}
\end{equation}
with
\begin{equation}
I(0) \,=\, I_b \, \exp[b(r_b/r_{e})^{1/n}].
\end{equation} 
The integrated luminosity out to a given radius for this model is
given by:
\begin{equation}
L(r) \, = \, \frac{2\pi n}{b^{2n}} \, r_e^2 \, I(0) \, \gamma(2n,b(r/r_e)^{1/n}),
\end{equation}
with $\gamma(a,x)$ being the incomplete gamma function.  We can now
determine the effective radius $r_{es}$ for Eqn.~\ref{sersextr} using
the effective radius equation:
\begin{equation}
2 \, L(r_{es}) = L(\infty),
\end{equation}
with $L(\infty)$ being the total luminosity.  For Eqn.~\ref{sersextr},
the effective radius equation becomes:
\begin{equation}
2 \, \gamma(2n,b(r_{es}/r_e)^{1/n}) \, = \, \Gamma(2n),
\end{equation}
where $\Gamma(a)$ is the complete gamma function.
On the other hand, if we have a pure S\'ersic law described by the
index $n$ the above equation is written as:
\begin{equation}
2 \, \gamma(2n,b_n) = \Gamma(2n).
\label{eq:sersicb}
\end{equation}
It follows immediately that:
\begin{equation}
b_n = b(r_{es}/r_e)^{1/n},
\end{equation}
or, equivalently,
\begin{equation}
b_n(1/r_{es})^{1/n} = b(1/r_e)^{1/n},
\label{identity}
\end{equation}
as we wanted to show.

\subsection{The Evaluation of $b$ for the \Bomba{} Model}

The quantity $b$ is used in the S\'ersic and \bomba{} models in order
to give $r_e$ the meaning of effective radius.  In order to evaluate
$b$, it is thus necessary to solve the implicit equation $2 L(r_e) =
L_T$.  For the S\'ersic profile ($b = b_n$), as is known, this
produces Eqn.~\ref{eq:sersicb}, given above.  For the \bomba{} model, $b$ 
is a function of the various parameters ($\alpha$, $\gamma$, $r_{b}$, 
and $r_{e}$) in the \bomba{} model, and can be determined by solving 
the following relation:
\begin{eqnarray}
2 \int_{b(r_b/r_e)^{1/n}}^{b(1/r_e)^{1/n}(r_b^\alpha +
r_e^\alpha)^{1/(n\alpha)}} e^{-x} x^{n(\gamma + \alpha) - 1}
(x^{n\alpha} - (b^n r_b/r_e)^{\alpha})^{(2-\gamma-\alpha)/\alpha} dx \; = \nonumber\\
\int_{b(r_b/r_e)^{1/n}}^{+\infty} e^{-x} x^{n(\gamma + \alpha) - 1}
(x^{n\alpha} - (b^n r_b/r_e)^{\alpha})^{(2-\gamma-\alpha)/\alpha} dx.
\end{eqnarray}
This assumes that $\alpha > 0$.  As $r_b \rightarrow 0$, we recover
the S\'ersic expression.  In the particular case $\alpha \rightarrow
\infty$ (sharp transition between inner power-law and outer S\'ersic
regimes), the equation simplifies to
\begin{equation}
\frac{1}{2 - \gamma}\bigg(\frac{r_b}{r_e}\bigg)^2 = \,
\frac{n}{b^{2n}} e^{b(r_b/r_e)^{1/n}}\left\{\Gamma(2n) \,+\,
\gamma(2n,b(r_b/r_e)^{1/n}) \,-\, 2\gamma(2n,b)\right\}.
\end{equation}

In practice, as long as $r_{b} \ll r_{e}$ and $\gamma < $1, the above
equation can be simplified even more:
\begin{equation}
\Gamma(2n) \, + \,
\gamma(2n,b(r_b/r_e)^{1/n}) \, \approx \, 2\, \gamma(2n,b).
\end{equation}

\subsection{Local Logarithmic Slope $\gammap$}

\citet{rest01} introduced $\gammap$ as a measure of the
(logarithmic) gradient of the luminosity profile at some specific
radius $r^{\prime}$:
\begin{equation}
\gammap \equiv - \bigg[\frac{d \log I}{d \log r}\bigg]_{r'}.
\end{equation}
For the Nuker law, \gammap{} is \citep[e.g.,][ Eqn.~8]{rest01}:
\begin{equation}
\gammap  = \frac{\gamma + \beta(r'/r_b)^\alpha}{1 + (r'/r_b)^\alpha}.
\end{equation}
As Rest et al.\ noted, this is a more accurate description of the
local logarithmic slope than the Nuker-law parameter $\gamma$ when the
transition between the two power-law regimes is soft (i.e, small
$\alpha$).  For the S\'ersic profile we have:
\begin{equation}
\gammap  = \frac{b}{n}\bigg(\frac{r'}{r_e}\bigg)^{1/n}.
\label{sersicgamma}
\end{equation}
Finally, for the \bomba{} model:
\begin{equation}
\gammap  = \frac{b}{n}\bigg(\frac{1}{r_e}\bigg)^{1/n}
r'^{\alpha}(r'^{\alpha}+r_b^{\alpha})^{1/(n\alpha)-1}+
\frac{\gamma(r_b/r')^{\alpha}}{1+(r_b/r')^{\alpha}}.
\end{equation}
As $r_b \rightarrow 0$, we recover the S\'ersic expression.  As
$\alpha \rightarrow \infty$, \gammap{} is described by the S\'ersic
value outside $r_{b}$ and is $= \gamma$ inside.

\subsection{Total Luminosity}

We assume the object is circular.  If the galaxy is elliptical the
following expressions must be multiplied by $b/a$, where $a$ and $b$
are semi-major and semi-minor axes, respectively.  The total
luminosity is defined as:
\begin{equation}
L_T = \int_0^{2\pi}\int_0^{+\infty} I(r) \, r \, dr \, d\theta.
\end{equation}

For a S\'ersic profile the total luminosity is then
\begin{equation}
L_T \, = \, \frac{2\pi n}{b^{2n}} \, \Gamma(2n)\, I(0)\, r_e^2,
\end{equation}
while for the \bomba{} model it is
\begin{equation}
L_T \, = \, 2\pi \, I' \, n\bigg(\frac{r_e}{b^n}\bigg)^{2}  
\int_{b(r_b/r_e)^{1/n}}^{+\infty}e^{-x}x^{n(\gamma + \alpha) - 1}
(x^{n\alpha} - (b^nr_b/r_e)^{\alpha})^{(2-\gamma-\alpha)/\alpha} \, dx.
\end{equation}
This expression is valid for $\alpha > 0$.  As $r_b \rightarrow
0$, we recover the S\'ersic expression.  In
the particular case $\alpha \rightarrow \infty$, this expression
becomes:
\begin{equation}
L_T \, = \, 2\pi
I_b \, \left\{\frac{r_b^2}{2-\gamma} \,+\, e^{b(r_b/r_e)^{1/n}}n\frac{r_e^2}{b^{2n}}
\bigg[\Gamma(2n)-\gamma(2n,b(r_b/r_e)^{1/n})\bigg]\right\}.
\end{equation}

\section{Contour Maps and Ellipse Fits} 
\label{app:figs}

In Figure~\ref{fig:efits} we display the isophotal contour maps and
ellipse fits for the WFPC2 mosaics of each of the galaxies we
analyzed.  Details of the data reduction can be found in
Section~\ref{sec:reduction}.

\section{Galaxies Rejected as Probable S0} 
\label{app:rejected}

The following galaxies met our selection criteria for size and for the
existence of WFPC2 archival images in the appropriate filters, but
were judged to have significant disks and thus be possible S0
galaxies, despite their formal classification as ellipticals.  We err
on the conservative side by considering the presence of bars and rings
to be evidence for an S0 galaxy; evidence for a bar includes the
appearance of the isophotes, peaks in ellipticity and accompanying
position-angle twists in the ellipse fits, and typical bar appearance
in unsharp masks \citep[see, e.g.,][]{erwin-sparke03}.  We also use
evidence from our attempts to fit the extra-nuclear ($r > 1\arcsec$)
light profiles (derived from the mosaic images) with both pure
S\'ersic and disk + bulge models: i.e., there are some galaxies for
which S\'ersic + exponential is clearly a better fit than pure
S\'ersic.

\textbf{NGC 596:} Source: \citet{lauer95}.  \citet{nieto92} argued 
that this was actually an SB0 galaxy; \citet{faber97}  also note that 
this galaxy has ``an S0-like outer envelope.''  Our fits to the light 
profile also suggest a disk + bulge morphology.

\textbf{NGC 2592:} Source: \citet{rest01}.  Kinematic evidence from
\nocite{rix98}Rix, Carollo, \& Freeman (1998) strongly suggests this
is an S0 galaxy; in addition, there is evidence for a bar in the PC
isophotes and unsharp masks.

\textbf{NGC 2699:} Source: \citet{rest01}.  Kinematic evidence from
\citet{rix98} strongly suggests this is an S0 galaxy; in addition,
there is clear evidence of a bar in the PC image (\nocite{rest01}Rest
et al.\ 2001 pointed to this galaxy as a providing a good example of
a misaligned inner structure, e.g., a bar).

\textbf{NGC 2778:} Source: \citet{rest01}.  Kinematic evidence from
\citet{rix98} strongly suggests this is an S0 galaxy; in addition,
there is good evidence for a bar in the PC image.  Analysis of the
light profile in \citet{kent85} and \citet{erwin03-smbh} also supports
an S0 (i.e., bulge + outer disk) interpretation.

\textbf{NGC 3608:} Source: \citet{lauer95}.  The light profile is 
significantly better fit with a disk + bulge model than by a pure 
S\'ersic model; see \citet{erwin03-smbh}.

\textbf{NGC 4121:} Source: \citet{rest01}.  There is clear evidence
for a bar in the PC image (``misaligned inner structure'' in
\nocite{rest01}Rest et al.), and the extra-nuclear light profile is
much better fit with a composite (bulge + disk) model than by a single
S\'ersic component.

\textbf{NGC 4564:} Source: \citet{rest01}.  Unsharp masking of the PC
image indicates that the elliptical feature dominating the isophotes
is a stellar ring, which we judge to be a signature of a significant
disk; there is some evidence for a nuclear bar as well.  Analysis of
the light profile in \citet{erwin03-smbh} also supports an S0 (i.e.,
bulge + outer disk) interpretation.

\textbf{NGC 4648:} Source: \citet{rest01}.  A very clear, strong bar
dominates the inner isophotes of the PC image (``misaligned inner
structure'' in \nocite{rest01}Rest et al.).

\textbf{NGC 5812:} Source: \citet{rest01}.  The light profile is
somewhat better fit with a disk + bulge model than by a pure S\'ersic
model; there is also weak evidence for a possible bar or ring in
the $r \approx 2$--5\arcsec{} isophotes.  This is probably the most 
uncertain ``S0'' classification in our rejected set.

\textbf{NGC 5813:} Source: \citet{rest01}.  The ellipticity steadily
increases outwards in this galaxy, from $\sim 0.1$ near the center to
$\sim 0.3$ at large radii, which is possible evidence for an outer
disk.  Analysis of the light profile indicates a disk + bulge 
structure as well.

\end{document}